\documentclass[12pt]{iopart}
\usepackage{color}

\usepackage{graphicx}  

\def\beqra{\begin{eqnarray}}
\def\eeqra{\end{eqnarray}}
\def\beq{\begin{equation}}
\def\eeq{\end{equation}}

\def\L{\Lambda}
\def\l{\lambda}

\def\bk{{\bf k}}
\def\vp{\varphi}

\def\bx{{\bf{x}}}
\def\bp{{\bf{p}}}
\def\bq{{\bf{q}}}

\def\half{\mbox{\small$\frac{1}{2}$}}

\def\agt{\stackrel{>}{\sim}}
\def\alt{\stackrel{<}{\sim}}
\begin{document}

\title{Resumming Cosmic Perturbations}

\author{Sabino Matarrese}
\address{Dipartimento di Fisica ``G. Galilei'', Universit\`a di Padova and 
INFN, Sezione di Padova, via Marzolo 8, I-35131, Padova, Italy}
\ead{sabino.matarrese@pd.infn.it}
\author{Massimo Pietroni}
\address{INFN, Sezione di Padova, via Marzolo 8, I-35131, Padova, Italy}
\ead{massimo.pietroni@pd.infn.it}

\begin{abstract}
Renormalization Group (RG) techniques have been successfully employed in 
quantum field theory and statistical physics. Here we apply RG methods
to study the non-linear stages of structure formation in the Universe. 
Exact equations for the power-spectrum, the bispectrum, and all higher 
order correlation functions can be derived for any underlying cosmological 
model. A remarkable feature of the RG flow is the emergence of an intrinsic 
UV cutoff, due to dark matter velocity dispersion, which improves the 
convergence of the equations at small scales. 
As a consequence, the method is able to follow the non-linear evolution of 
the power-spectrum 
down to zero redshift and to length-scales where perturbation theory 
fails. Our predictions accurately fit the results of $N$-body simulations 
in reproducing the ``Baryon Acoustic Oscillations'' features of 
the power-spectrum, which will be accurately measured in future galaxy 
surveys and will provide a probe to distinguish among different dark energy 
models.
\end{abstract}

\maketitle
\section{Introduction}

Precision Cosmology relies on the capability of accurately extracting 
cosmological parameters from measurements of observables related to 
Cosmic Microwave Background (CMB) temperature anisotropy and polarization 
as well as to the Large-Scale Structure (LSS) of the Universe. 
In the CMB context, the tool allowing a comparison of data and theory, 
at the percent level, is perturbation theory (PT).
One would however like to reach the same level of precision also in the 
study of the other main cosmological probe, the LSS of the Universe. 
In particular, the location and amplitude of Baryon Acoustic Oscillations 
(BAO), wiggles in the matter power-spectrum produced by the coupling 
of baryons to radiation by Thomson scattering in the early universe, 
for wavenumbers in the range $k\simeq 0.05 - 0.3 \;h \mathrm{Mpc}^{-1}$, 
have the potential to constrain the expansion 
history of the Universe and the nature of the Dark Energy \cite{useBAO}. 
BAO's have recently been detected both in the 2dF and 
SDSS surveys data \cite{BAOobs}, and are going to be measured in the 
near future in a series of high-redshift surveys \cite{BAOhigz}.

Due to the higher degree of non-linearity of the underlying density 
fluctuations compared to those relevant for the CMB, accurate computations 
of thepower-spectrumin the BAO region is challenging, and has so far been 
best afforded by means of high-resolution $N$-body simulations (see, e.g.
Ref.~\cite{White}).
In particular, the study of BAO to constrain Dark Energy models 
requires a precision of a few percent in the theoretical predictions 
for the matterpower-spectrumin the relevant wavenumber range \cite{Komatsu}. 
Including higher orders in PT \cite{PT} is known to give a poor 
performance in this range, while all available fitting functions 
for the non-linearpower-spectrum(e.g. refs.~\cite{Peacock,Smith}) are uncapable to 
reach the required level of accuracy \cite{White}, leaving N-body simulations 
as the only viable approach to the problem. 

Recently, however, PT has experienced a 
renewed interest, mainly motivated by two reasons. First, next generation
galaxy surveys \cite{BAOhigz} are going to measure thepower-spectrumat 
large redshift, where the fluctuations are still in the linear regime and 
1-loop PT is expected to work \cite{Komatsu}. Second,
Crocce and Scoccimarro \cite{Crocce1, Crocce2} have shown that 
the perturbative 
expansion can be reorganized in a very convenient way, which allows the 
use of standard tools of field theory, like Feynman diagrams. They 
managed to compute the two-point correlator between density or velocity 
field fluctuations at different times (the `propagator') by  resumming an 
infinite class of diagrams at all orders in PT. Other approaches 
can be found in Ref.~\cite{RGLSS}. 

In this paper, we will present a new approach to the problem, based on 
Renormalization Group (RG) 
techniques developed in quantum field theory and 
statistical physics \cite{RGreview}. We will give the full
details of the formulation of the method and of its application to the 
computation of the power-spectrum 
in the BAO region, whose results have been recently presented in 
Ref.~\cite{ms1}.

RG methods are particularly suited to physical situations in 
which there is a separation between the scale where one is supposed to 
control the `fundamental' theory and the scale were measurements are 
actually made. Starting from the fundamental scale, the RG flow 
describes the gradual inclusion of fluctuations at scales closer and 
closer to the one relevant to measurements. The new fluctuations 
which are included at an intermediate step, feel an effective theory,
which has been `dressed' by the fluctuations already included. 
In the present case, the RG flow will start from small wavenumbers 
$k$, where linear theory works, to reach higher and higher $k$. 

One of the most important results of our analysis is that an intrinsic  
ultra-violet (UV) cutoff emerges from the RG flow, which suggest that 
the non-linear gravitational evolution itself implies a small-scale smoothing, 
so that our description can work also on fairly non-linear 
scales. The main result of this paper is the 
evaluation of the non-linear matter power-spectrum in a standard $\Lambda$CDM
model, including a very accurate fit of the BAO features obtained in 
$N$-body simulations, down to zero redhift and to length-scales 
where any previous analytical or semi-analytical techniques have failed. 
Moreover, our approach allows a straightforward extension to more general 
cosmologies with dynamical dark energy. 

The plan of the paper is as follows. In Section 2 we introduce the 
equations that govern the dynamics of self-gravitating dark matter 
in the fluid limit and, following Ref.~\cite{Crocce1}, 
we recast them in the compact form of a single non-linear equation, 
for a suitable field doublet in Fourier space. In Section 3 we obtain the 
generating functional of n-point correlation functions for our system in the form of a path-integral 
which incorporates the statistics of initial conditions (here assumed 
to be Gaussian, for simplicity). The general formulation of the RG equations is given in Section 4. In Sects. 5 and 6 we derive the equations and present our solutions for the non-linear 
propagator and PS, respectively. 
Conclusions and future prospects are discussed in Sect. 7. 

\section{Eulerian theory revisited}
The distribution of a gas of DM particles of mass $m$ is described 
by the function $f(\bx,\,\bp, \,\tau)$, 
where $\bx$ represents the comoving spatial coordinate, $\bp$ is 
given by \[{\bp}=a m \frac{d {\bf x}}{d \tau}\,,\]
($a$ being the scale factor) and $\tau$ is the conformal time.

The evolution of $f$ is governed by the collisionless Boltzmann 
equation, a.k.a. the Vlasov equation,
\beq\frac{\partial f}{\partial \tau}+\frac{\bf p}{a m}\cdot 
\nabla f - a m \nabla \phi \cdot \nabla_{\bf p} f=0\,.\eeq
$\phi$ is the gravitational potential which, on subhorizon scales, 
obeys  the Poisson equation,
\beq\nabla^2 \phi = \frac{3}{2}\,\,{\cal H}^2  \,\delta\,,\eeq
with ${\cal H}= d \log a/d \tau$ and $\delta$ the mass-density 
fluctuation, 
\beq \int d^3{\bf p} \,f({\bf x}, {\bf p}, \tau) \equiv \rho({\bf x}, 
\tau)\equiv \overline{\rho}(\tau) [1+\delta({\bf x},
\tau)]\,,
\label{mom0}
\eeq
 having assumed an Einstein-de Sitter background cosmology.
 
A full solution of the --non-linear and non-local--  Vlasov equation is a 
formidable task. A more affordable
strategy is to take {\it moments} of the distribution functions. 
Along with the zeroth order one, defined in 
Eq.~(\ref{mom0}), one can define the infinite tower of higher order 
moments,
\beqra
&&\int d^3{\bf p} \,\frac{p_i}{am}f({\bf x}, {\bf p}, \tau) \equiv 
\rho({\bf x}, \tau) {v_i}({\bf x}, \tau)  \nonumber\\
&&\int d^3{\bf p} \,\frac{ p_i \, p_j}{a^2 m^2}f({\bf x}, {\bf p}, \tau) 
\equiv \rho({\bf x}, \tau) { v}_i({\bf x}, \tau){ v}_j
({\bf x}, \tau) + \sigma_{ij}({\bf x}, \tau) \nonumber\\
&& \qquad\qquad\qquad\qquad\qquad \ldots\,\qquad\qquad\qquad\qquad\qquad\qquad.
\eeqra

The definitions above, once inserted in the Vlasov equation, 
give rise to an infinite hierarchy of equations involving moments 
of higher and higher order \cite{PT}. The system can be truncated by 
setting to zero the stress tensor $\sigma_{ij}$. Although the latter term 
would arise as soon as orbit-crossing appears in the non-linear 
evolution of our collisionless system, this procedure, 
dubbed `single stream approximation', 
is formally self-consistent, since no non-vanishing $\sigma_{ij}$ is generated 
by the above set of moment equations once it is set to zero in the initial 
conditions. Moreover, as we will see in the following, an intrinsic  
UV cutoff emerges from the RG flow.

In this case, one is left with just two equations
\beqra
&&\frac{\partial\,\delta}{\partial\,\tau}+
{\bf \nabla}\cdot\left[(1+\delta) {\bf v} \right]=0\,,\nonumber\\
&& \frac{\partial\,{\bf v}}{\partial\,\tau}+{\cal H} {\bf v} + ( {\bf v} 
\cdot {\bf \nabla})  {\bf v}= - {\bf \nabla} \phi\,,
\label{Euler}
\eeqra
which are the continuity and Euler equations, respectively.

Defining, as usual, the velocity divergence $\theta(\bx,\,\tau) = 
\nabla \cdot {\bf v} (\bx, \,\tau)$, and going to
Fourier space, Eqs.~(\ref{Euler}) read
\beqra
&&\frac{\partial\,\delta({\bf k}, \tau)}{\partial\,\tau}+\theta({\bf k}, 
\tau) \nonumber\\
&& \qquad + \int d^3\bq\, d^3\bp \,\delta_D({\bf k}-\bq-\bp)
 \alpha(\bq,\bp)\theta(\bq, \tau)\delta(\bp, \tau)=0\,,\nonumber\\
&&\frac{\partial\,\theta({\bf k}, \tau)}{\partial\,\tau}+
{\cal H}\,\theta({\bf k}, \tau)  +\frac{3}{2} {\cal H}^2 
\delta({\bf k}, \tau)\nonumber\\
&& \qquad   +\int d^3\bq \,d^3\bp \,\delta_D({\bf k}-\bq-\bp) 
\beta(\bq,\bp)\theta(\bq, \tau)\theta(\bp, \tau) = 0 \,.\label{EulerFourier}
\eeqra
The non-linearity and non-locality of the Vlasov equation survive in 
the two functions
\beq\alpha(\bp,\bq )= \frac{(\bp + \bq) \cdot \bp}{p^2}\,,\quad \quad
\beta(\bp,\bq ) = \frac{(\bp + \bq)^2 \,
\bp \cdot \bq}{2 \,p ^2 q^2}\,,
\eeq
which couple different modes of density and velocity fluctuations. 

Setting $\alpha(\bp,\bq )=\beta(\bp,\bq )=0$, the equations can be solved as 
\beqra
&& \delta({\bf k},\,\tau) = \delta({\bf k},\,\tau_i) 
\left(\frac{a(\tau)}{a(\tau_i)}\right)^m\,,\nonumber\\
&&-\frac{\theta({\bf k},\,\tau)}{\cal H}= m\, \delta({\bf k},\,\tau)\,, 
\qquad\qquad \quad (m = 1,\,-3/2),
\eeqra
where we recognize the growing ($m=1$) and decaying ($m=-3/2$) 
modes of the linear order equations of standard cosmological 
PT in the Newtonian limit.

Following Crocce and Scoccimarro \cite{Crocce1, Crocce2} one 
can write Eqs.~(\ref{EulerFourier}) in a compact form. First, 
we introduce the doublet $\vp_a$ ($a=1,2$), given by
\beq\left(\begin{array}{c}
\varphi_1 ( {\bf k}, \eta)\\
\varphi_2 ( {\bf k}, \eta)  
\end{array}\right)
\equiv 
e^{-\eta} \left( \begin{array}{c}
\delta  ( {\bf k}, \eta) \\
-\theta  ( {\bf k}, \eta)/{\cal H}
\end{array}
\right)\,,
\label{doppietto}
\eeq
where the time variable has been replaced by the logarithm of 
the scale factor,
\[ 
\eta= \log\frac{a}{a_{in}}\,,
\]
$a_{in}$ being the scale factor at a conveniently remote epoch, 
were all the relevant scales are well inside the linear regime. 
Notice that, compared with the definition used by Crocce and 
Scoccimarro, we have a $e^{-\eta}$ factor overall, such that the 
linear growing mode corresponds to $\vp_a = \mathrm{const}$.

Then, we define a {\it vertex} function, 
$\gamma_{abc}({\bf k},{\bf p},{\bf q}) $ ($a,b,c,=1,2$) 
whose only non-vanishing elements are
\beqra
&&\gamma_{121}({\bf k},\,{\bf p},\,{\bf q}) = 
\frac{1}{2} \,\delta_D ({\bf k}+{\bf p}+{\bf q})\, 
\alpha(\bp,\bq)\,,\nonumber\\
&&\gamma_{222}({\bf k},\,{\bf p},\,{\bf q}) = 
\delta_D ({\bf k}+{\bf p}+{\bf q})\, \beta(\bp,\bq)\,,
\label{vertice}
\eeqra
and 
$\gamma_{121}({\bf k},\,{\bf p},\,{\bf q})  = 
\gamma_{112}({\bf k},\,{\bf q},\,{\bf p}) $.

The two equations (\ref{EulerFourier}) can now be rewritten as 
\beq
\left(\delta_{ab} \partial_\eta+\Omega_{ab}\right) 
\varphi_b({\bf k}, \eta) =  e^\eta 
\gamma_{abc}({\bf k},\,-{\bf p},\,-{\bf q})  
\varphi_b({\bf p}, \eta )\,\varphi_c({\bf q}, \eta )\,,
\label{compact}
\eeq
where 
\[{\bf \Omega} = \left(\begin{array}{cc}
1 & -1\\
-3/2 & 3/2
\end{array}
\right)\,,\]
and repeated indices/momenta are summed/integrated over.
Besides the vertex, the other building block of the perturbation theory 
we are going to use is the linear retarded {\it propagator}, 
defined as the operator giving the evolution of the field $\vp_a$ 
from $\eta_b$ to $\eta_a$,
\beq
\varphi_a^0( {\bf k},\eta_a) = g_{ab}(\eta_a,\eta_b)  
\varphi_b^0({\bf k},\eta_b)\,,\quad\quad(\eta_a>\eta_b)
\label{linearsol}
\eeq
where the subscript $``0"$ indicates solutions of the linear 
equations (obtained in the  $e^\eta\,\gamma_{abc} \to 0$ limit).

The propagator $g_{ab}(\eta_a,\eta_b)$ can be explicitly 
computed by solving the equation
\beq\left(\delta_{ab} \partial_{\eta_a}+\Omega_{ab}\right) 
g_{bc}(\eta_a, \eta_b) = \delta_{ac}\, \delta_D(\eta_a-\eta_b)\,,
\label{green}
\eeq
with causal boundary conditions (see \cite{Crocce1}), getting,
\beq 
g_{ab}(\eta_a,\eta_b) =\left[ {\bf B} + {\bf A}\, e^{-5/2 
(\eta_a -\eta_b)}\right]_{ab}\, \theta(\eta_a-\eta_b)\,,
\label{proplin}
\eeq
 with $\theta$ the 
step-function, and
\beq {\bf B} = \frac{1}{5}\left(\begin{array}{cc}
3 & 2\\
3 & 2
\end{array}\right)\,\qquad {\mathrm{and}} \qquad
{\bf A} = \frac{1}{5}\left(\begin{array}{rr}
2 & -2\\
-3 & 3
\end{array}\right)\,.
\eeq
The growing ($\vp_a \propto \mathrm{const.}$) and the decaying 
($\vp _a\propto \exp(-5/2 \eta_a)$) modes can be selected by 
considering initial fields $\vp_a$ proportional to 
\beq u_a = \left(\begin{array}{c} 1\\ 
1\end{array}\right)\,\qquad\mathrm{and} 
\qquad v_a=\left(\begin{array}{c} 1\\ -3/2\end{array}\right)\,,
\label{ic}
\eeq
respectively.

To extend the validity of this approach to $\Lambda$CDM, we will
reinterpret the variable $\eta$ as the logarithm of the linear
growth factor of the growing mode, i.e. 
$\eta=\ln (D^+/D^+_{in})$. This approximation has been shown to 
accurately fit N-body simulations for different cosmologies 
(e.g. Refs.~\cite{Komatsu,Nusser}). A closer look at the problem shows 
that this approximation relies on the ansatz of setting artificially to one 
the quantity $\e(\Omega_m) \equiv \Omega_m /f^2$, with $f=d \ln D_+/d\ln 
a$, which is anyway very close to unity for most of the history 
of the Universe in a wide class of cosmologies.
Moreover, as shown in Ref.~\cite{Crocce2}, the 
slight deviation of $e(\Omega_m)$ from unity would only affect the decaying 
mode, thus making the above approximation extremely accurate.  

\section{Generating functionals} 
The aim of this section is to apply methods familiar in 
quantum field theory to construct generating functionals for 
quantities like the power-spectrum, bispectrum, propagator and any 
other object of interest. 
The starting point is to write down an action giving the 
equation of motion (\ref{compact}) at its extrema. One 
can realize that a new, auxiliary, doublet field  $\chi_a$ has 
to be introduced to this aim, and that the action is given by 
\beqra
 \label{action}
S&=& \int d\eta \left[ \chi_a(-\bk,\eta)\left(\delta_{ab} 
\partial_\eta+\Omega_{ab}\right) \varphi_b( {\bf k},\eta)\right.\nonumber\\
&& -\left. e^\eta \gamma_{abc}(-\bk, -\bp,-\bq)
\chi_a(\bk,\eta)\varphi_b(\bp,\eta)
\varphi_c( \bq ,\eta) \right] \,.
\eeqra

The introduction of the auxiliary field $\chi_a$ is required by 
the bilinear term being first order in the `time' derivative $\partial_\eta$. 
Indeed, a term of the form $\vp_a\partial_\eta\vp_a$ would vanish 
upon integration by parts. 
Using Eq.~(\ref{green}), the bilinear part in the first line 
of the action can be also written as
\beq
S_2=\int d\eta_a\, d\eta_b \,\chi_a(-{\bk},\eta_a) g_{ab}^{-1} 
(\eta_a,\eta_b) \varphi_b( {\bf k}, \eta_b)\,.
\eeq
Varying the action (\ref{action}) with respect to $\chi_a$ gives 
precisely Eq.~(\ref{compact}), while varying with respect to $\vp_a$ gives
 an equation solved by $\chi_a=0$. The r\^{o}le of the $\chi_a$ field
is not merely that of allowing to write the action above, but more
physically, it is related to the statistics of initial conditions, as 
we will see below. In the following, in order to simplify the notation, 
we will omit momentum dependence when it is obvious.

Being the system classical, the probability of having a field 
$ \vp_a(\eta_f)$ at time $\eta_f$, starting with an initial condition 
$\vp_a(0)$ is a (functional) delta function:
\beq
P[ \vp_a(\eta_f);\, \vp_a(0)] = \delta\left[\vp_a(\eta_f) - 
\bar{\vp}_a[\eta_f;\,\vp_a(0)]\right]\,,
\label{prob}
\eeq
where $ \bar{\vp}_a[\eta_f;\,\vp_a(0)]$ is the solution to the 
equation of motion (\ref{compact}) with initial condition $\vp_a(0)$. 
Using the action (\ref{action}), where the $\chi_a$ field enters 
linearly, the delta function can be given a path integral representation, 
\beq
P[ \vp_a(\eta_f);\, \vp_a(0)] = {\cal N} \int {\cal D}''\vp_a\,{\cal D} 
\chi_b \,e^{i S}\,,
\eeq
where the double prime on the measure for $\vp_a$ means that it is 
kept fixed at the two extrema $\eta=0$ and $\eta=\eta_f$. 
In the following, the field-independent normalization 
${\cal N}$ will be set to unity.

We then define a generating functional 
by following the standard procedure, {\it i.e.} by 
introducing sources for $\vp_a$ 
and $\chi_b$ and by summing over all the possible final states, {\it i.e.},
\beqra
&&Z[J_a,\, K_b;\, \vp_a(0)] \equiv \int {\cal D}\vp_a(\eta_f)  
\int {\cal D}''  \vp_a {\cal D}\, \chi_b\,\times \nonumber\\
&&\exp\left\{i\int_0^{\eta_f} d\eta \, \chi_a(\delta_{ab}\partial_\eta+
\Omega_{ab})\vp_b- e^\eta\, \gamma_{abc}\chi_a\vp_b\vp_c  + J_a\vp_a +
 K_a \chi_a
\right\}.
\label{Zp}
\eeqra
Finally, since we are interested in statistical systems, we average the 
probabilities over the initial conditions with a statistical weight function
 for the physical fields $\vp_a(0)$,
\beq
Z[J_a,\, K_b;\,C's] = \int {\cal D}\vp_a(0)\, W[\vp_a(0),\,C's]\,
 Z[J_a,\, K_b;\, \vp_a(0)] \,.
\label{fullZ}
\eeq
In general, the initial weight can be expressed in terms of the 
initial $n$-point correlations as \cite{Calzetta} 
\beqra
&& W[\vp_a(0),\,C's]= \exp \left\{- \vp_a(\bk,\,0)C_a(\bk) - 
\vp_a(\bk_1,\,0) C_{ab}(\bk_1,\,\bk_2) \vp_b(\bk_2,\,0)\right. \nonumber \\
&&\left. +\vp_a(\bk_1,\,0) \vp_b(\bk_2,\,0) \vp_c(\bk_3,\,0) 
C_{abc}(\bk_1,\,\bk_2,\,\bk_3)  +\cdots \right\}\,.
\eeqra
In the case of Gaussian initial conditions, the only non-zero 
initial correlation is the quadratic one, and the
weight function reduces to the form
\beq 
 W[\vp_a(0),\,C_{ab}]=
\exp{\left\{-\half \vp_a(\bk,\,0) C_{ab}(k) \vp_b(-\bk,\,0)\right\}}\,,
\label{Wgauss}
\eeq
where 
\beq {\bf C^{-1}}_{ab}(k)=  P^0_{ab}(k) \equiv w_a w_b P^0(k)\,,
\eeq
with $P^0(k)$ the initial power-spectrum and the two-component 
vector $ w_a$ is a combination of $u_a$ and $ v_a$ in Eq.~(\ref{ic}) 
describing the initial mixture of growing  and decaying modes \cite{Crocce1}.
In the following, we will restrict the initial conditions to the 
Gaussian case, Eq.~(\ref{Wgauss}), leaving the 
study of the effect of primordial non-Gaussianity to future work.

The linear theory limit, $e^\eta \gamma_{abc}\to 0$, corresponds to
the tree-level of PT. 
In this limit the path integral can be explicitly computed. 
Performing first the $\chi_a$ integral, and then the 
$\int {\cal D}\vp_a(\eta_f)  {\cal D}''  \vp_a $ ones, we obtain \cite{Gozzi} 
 \beqra
&& Z_0[J_a,\, K_b;\,P^0] = \nonumber\\
&&\int {\cal D} \vp_a(0) \, \exp\left\{-\half \vp_a(\bk,\,0) 
\,({\bf {P^0}^{-1}})_{ab}(k) \,\vp_b(-\bk,\,0) +  
i \int_0^{\eta_f} d\eta\,J_a \tilde{\vp}_a \right\}\,,
\eeqra 
where $\tilde{\vp}_a$ is the solution of the classical equation of 
motion with source $K_a$,
\beq
\left( \delta_{ab} \partial_\eta+\Omega_{ab}\right)\tilde{\vp}_b(\eta)
 = - K_a(\eta)\,,
\eeq 
which is given by
\beq 
\tilde{\vp}_a(\eta_a) = { \vp}^0_a(\eta_a) -\int 
d\eta_b g_{ab}(\eta_a,\eta_b) K_b(\eta_b)\,,
\eeq
with ${ \vp}^0_a(\eta)$ the known sourceless zeroth-order solution, 
see Eq.~(\ref{linearsol}).

The remaining integral on the initial conditions $\vp_a(0)$ can 
also be performed, leading to the result
\beqra
&&Z_0[J_a,\, K_b;\,P^0]=\exp\left\{
-  \int d\eta_a d\eta_b \left[\half J_a(\bk,\eta_a) 
P^L_{ab}(k\,;\eta_a,\,\eta_b) J_b(-\bk,\eta_b)\right.
\right.\nonumber \\
&&\qquad\qquad\qquad\qquad \qquad\left. \left.+i J_a(\bk,\eta_a) 
g_{ab}(\eta_a,\eta_b) 
K_b(-\bk,\,\eta_b)\right]\right\}\,,
\label{zfree}
\eeqra
where $ P^L_{ab}$ is the power-spectrum evolved at linear order,
\beq
P^L_{ab}(k\,;\eta_a,\,\eta_b) = g_{ac}(\eta_a,0) g_{bd}(\eta_b,0) 
P^0_{cd}(k)\,.
\label{corrzero}
\eeq

Starting from this explicit expression we can recover the results of 
linear theory. For instance, the power-spectrum,
\beq
\langle  { \vp}_a(\bk, \eta_a) { \vp}_b(\bk^\prime, \eta_b)\rangle \equiv 
\delta(\bk+\bk') 
P_{ab}(k\,;\eta_a,\, \eta_b)\,,
\label{ps}
\eeq
is given (at linear order) by the double derivative of $Z_0$ with respect to 
the source $J_a$, 
\beq \frac{(-i)^2}{Z_0} \left. \frac{\delta^2 Z_0[J_a,\, 
K_b;\,P^0]}{\delta J_a(\bk,\eta_a)\,\delta J_b(\bk^\prime,
\eta_b)}\right|_{J_a,\,K_b=0}= \delta(\bk+\bk') 
P_{ab}^L(k\,;\eta_a,\,\eta_b) \,.\label{ps0}
\eeq
Using Eq.~(\ref{linearsol}) in (\ref{ps}) we recover 
the linear behavior of the power-spectrum, Eq.~(\ref{corrzero}).

The  cross-derivative gives the retarded propagator 
\beq 
\delta({\bk}+{\bk}') g_{ab}(\eta_a,\,\eta_b) 
=\frac{i}{Z_0} \left. \frac{\delta^2 Z_0[J_a,\, K_b;\,P^0]}{\delta 
J_a(\bk, \eta_a)\delta K_b(\bk^\prime, \eta_b)}\right|_{J_a,\,K_b=0} \,.
\label{gr}
\eeq

Thus, from a single object, $Z_0$, we are able to obtain all 
the quantities of interest, that is, the
propagator, the power-spectrum, and all higher order correlation
functions, by taking appropriate derivatives of it with respect to the
sources. The generalizations of Eqs.~(\ref{ps0}) and (\ref{gr}) to the
power-spectrum and propagator of the full non-linear theory will be
given in Eq.~(\ref{w2}) below.

\begin{figure}
\centerline{\includegraphics[width = 4in , 
keepaspectratio=true]{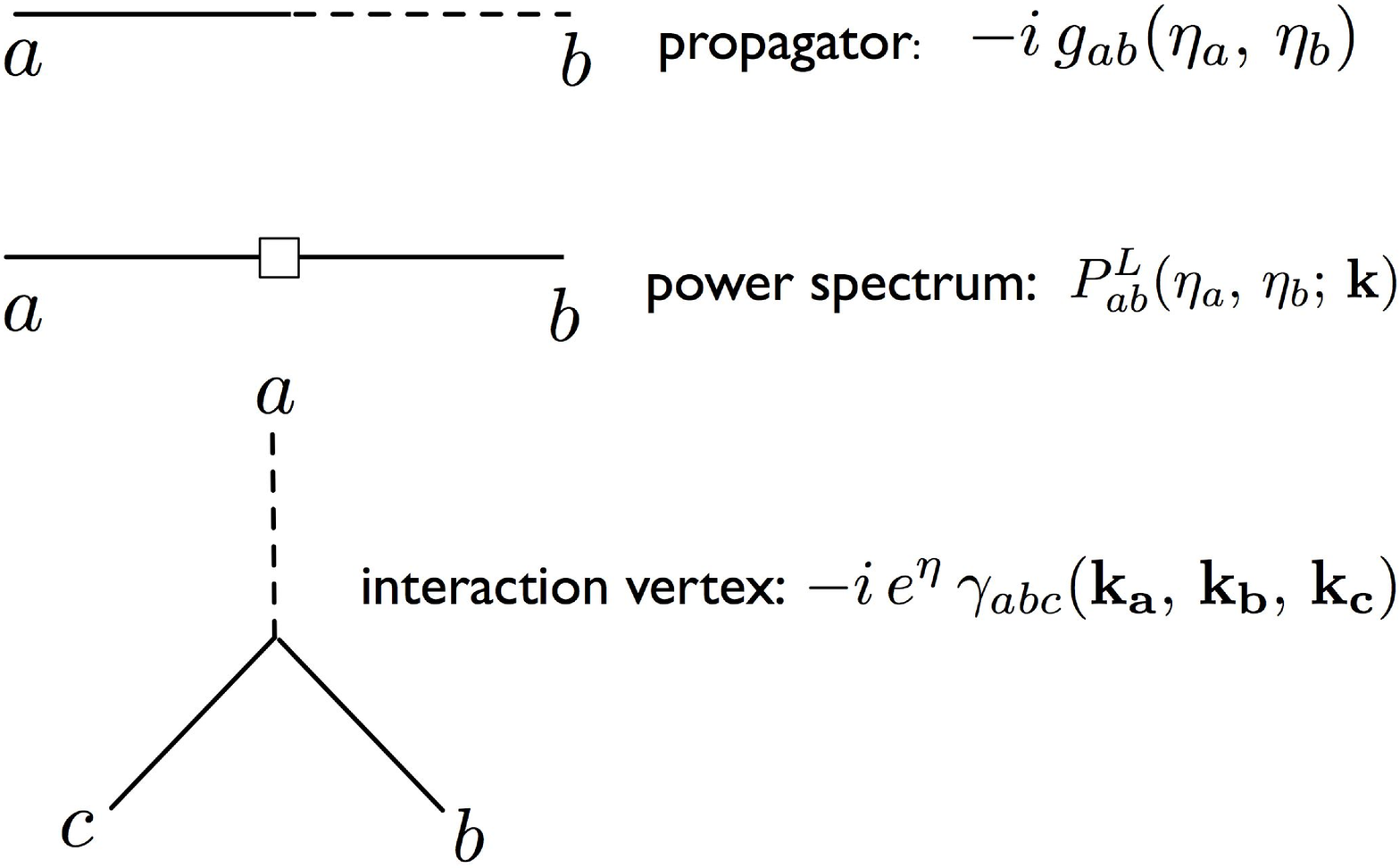}}
\caption{The Feynman rules.}
\label{Feynman}
\end{figure}

Turning the interaction $\gamma_{abc}$ on, the generating 
functional (\ref{fullZ}) can be rewritten as
\beqra
&& Z[J_a,\, K_b;\,P^0]=\nonumber\\
&&
 \exp\left\{-i\int d\eta\, e^\eta\,\gamma_{abc}\left( 
\frac{-i\delta}{\delta K_a}\frac{-i\delta}{\delta J_b}
 \frac{-i\delta}{\delta J_c}
\right)\right\} \, Z_0[J_a,\, K_b;\,P^0]\,, \label{fullZ2}
\eeqra
with $Z_0$ given by Eq.~(\ref{zfree}).
Higher orders in PT are obtained by 
expanding the exponential in powers of 
$\gamma_{abc}$. From this expression for $Z$ one can read out 
the Feynman rules. The three fundamental building blocks, 
{\it i.e.} the propagator $g_{ab}$, the linearly evolved 
power-spectrum $P^L_{ab}$, and the trilinear vertex 
$e^\eta\,\gamma_{abc}$,
can be represented by the Feynman diagrams in Fig.~\ref{Feynman}. 
Continuous and dashed lines
indicate $\vp_a$  and $\chi_a$ legs, respectively. 
Feynman diagrams constructed by these rules are 
in one to one correspondence with those considered in 
\cite{Crocce1,Crocce2}.

The expression (\ref{fullZ2}) is equivalent, at 
any order in PT, to 
\beqra
\label{genf}
Z[J_a,\, K_b;\,P^0] 
&=&\int {\cal D}  \vp_a {\cal D} \chi_b
\exp \biggl\{-\half \int d\eta_a d\eta_b   \chi_a P^0_{ab} 
\delta(\eta_a) \delta(\eta_b)\chi_b \nonumber\\
&+& 
i \int d\eta \left[ \chi_a g^{-1}_{ab} \vp_b  -  
e^\eta\,\gamma_{abc} \chi_a \vp_b \vp_c + 
J_a \vp_a + i K_b \chi_b\right]\biggr\} \,, 
\eeqra
which will be useful for the derivation of the RG equations.
Notice that the primordial power-spectrum, $P^0_{ab}$, 
is directly coupled to $\chi$-fields only, showing the r\^ole 
of these fields in encoding the information on the statistics of 
the initial conditions.

Proceeding on the standard path of field theory, we will also consider 
the generator of connected Green functions
\beq W=-i \log Z\,,
\label{WW}
\eeq
through which we can define expectation values of the fields $\vp_a$ 
and $\chi_b$ in the presence of 
sources,
\beq
\vp_a[J_c, K_d] =\frac{\delta W[J_c, K_d]}{\delta J_a}\,,\;\;\;\;\;
\chi_b[J_c, K_d]=\frac{\delta W[J_c, K_d]}{\delta K_b}\,.
\label{fields}
\eeq
When it is not ambiguous, we will use the same notation for the fields 
and their expectation values. Full, connected Green functions can be 
expressed in terms of full propagators and full one-particle irreducible 
(1PI) Green functions, see for instance Eqs.~(\ref{fullP1}), (\ref{fullP2}), 
(\ref{fullG}), and (\ref{W4}) below. 
Therefore it is useful to consider also the generating functional for 
1P1 Green functions, {\it i.e.} the effective action.
It is defined, as usual, as the Legendre 
transform of $W$,
\beq
\Gamma[\vp_a,\, \chi_b] =  W[J_a, K_b] - \int d\eta\,d^3 \bk \left(J_a \vp_a 
+ K_b \chi_b \right)\,.
\label{legendre}
\eeq
Derivatives of $\Gamma$ with respect to $\vp_a$ and $\chi_a$ give rise 
to 1PI
$n$-point functions. Notice that 1PI functions with all the external legs 
of type $\vp_a$ vanish at any order in PT, {\it i.e.} 
\beq
\left.\frac{\delta^n \Gamma[\vp_a,\chi_b]}{\delta \vp_{a_1}(\eta_1) 
\cdots \delta \vp_{a_n}(\eta_n)}\right|_{\vp_a=\chi_b=0}=0 \qquad\qquad
\mathrm{for \;any}\; n\,.
\label{jvp}
\eeq
Indeed, the contributions to a 1PI $n$-point function at $l$-loop 
order contain the product of $m=n+2(l-1)$  basic vertices,
 \beq  
 \chi \vp \vp(\eta_1)\;\chi \vp \vp(\eta_2)\cdots \chi \vp \vp(\eta_m)\;.
 \eeq
 In order to have a 1PI function, at most one of the fields in 
each vertex can be an `external' field, that is, it is not joined 
to fields in the other vertices via a
propagator ($\chi-\vp$ connection) or a power-spectrum ($\vp-\vp$ 
connection). Moreover, in order to have a $n$-point function like 
Eq.~(\ref{jvp}), with no $\chi$-field as an external field, every $\chi$ 
has to be contracted with a $\vp$ field belonging to a different vertex, 
via a retarded propagator. One can then realize that any diagram 
potentially contributing to (\ref{jvp}) contains at least one closed loop of 
propagators, which vanish due to the presence of the causal
$\theta$-functions in $\eta$.

Considering the second derivatives of the effective action, 
computed at $\vp_a=\chi_a=0$, we can then define
\beqra
&& \Gamma^{(2)}_{\vp_a \vp_b} =0 \,,\nonumber\\
&& \Gamma^{(2)}_{\vp_a \chi_b}\equiv g^{-1}_{ba} - \Sigma_{\vp_a \chi_b}
\,,\nonumber\\
&& \Gamma^{(2)}_{\chi_a \vp_b} \equiv g^{-1}_{ab} - 
\Sigma_{\chi_a \vp_b}\,,\nonumber\\
&&\Gamma^{(2)}_{\chi_a \chi_b}\equiv i  P^0_{ab}(k) \delta(\eta) \delta(\eta_b)
 + i \Phi_{ab}\,,
\label{G2}
\eeqra
where we have isolated the `free' ({\it i.e.} linear) parts, which 
can be read off from Eq.~(\ref{genf}), and the full, 1PI, two-point 
functions have been defined as
\beq
\delta(\bk_a+\bk_2)\,\Gamma^{(2)}_{\vp_a \vp_b} \equiv \left.
\frac{\delta^2 \Gamma[\vp_a,\chi_b]}{\delta \vp_a \delta \vp_b}
\right|_{\vp_a,\chi_b=0},
\eeq
and so on.

The full power-spectrum and propagators are given by second derivatives of $W$,
\beqra
&&\left. \frac{\delta^2 W}{\delta J_a\, \delta J_b}\right|_{J_a,\,K_b=0} 
\equiv i \delta({\bk}+{\bk}') P_{ab}\,,\nonumber\\
&&\left. \frac{\delta^2 W}{\delta J_a\, \delta K_b}\right|_{J_a,\,K_b=0} 
\equiv - \delta({\bk}+{\bk}') G_{ab} \,,\nonumber\\
&&\left. \frac{\delta^2 W}{\delta K_a\, \delta J_b}\right|_{J_a,\,K_b=0} 
\equiv - \delta({\bk}+{\bk}') G_{ba} \,,\nonumber\\
&&\left. \frac{\delta^2 W}{\delta K_a\, \delta K_b}\right|_{J_a,\,K_b=0} =0 \,.
\label{w2}
\eeqra
Using the definitions (\ref{fields}) and (\ref{legendre}) one can verify 
that the four quantities in Eq.~(\ref{w2}) form a matrix that is minus the 
inverse of that formed by the four quantities in (\ref{G2}), which implies 
that the $\delta^2 W/\delta K^2$ entry vanishes. We can then express the full
 propagator and power-spectrum in terms of the free ones and the 
`self-energies' appearing in eq~(\ref{G2}). Inverting (\ref{G2}) one then gets
\beq
P_{ab} = P^I_{ab}+P^{II}_{ab}\,,\label{fullP1}
\eeq
where
\beqra
\label{fullP2}
&&P^I_{ab}(k; \eta_a, \eta_b) = G_{ac}(k;\eta_a,0)
G_{bd}(k;\eta_b,0) P^0_{cd}(k)\,,\nonumber\\
&&P^{II}_{ab}(k; \eta_a, \eta_b)= \int_0^{\eta_a} d s_1
\int_0^{\eta_b} d s_2\,
G_{ac}(k;\eta_a,s_1)
G_{bd}(k;\eta_b, s_2) 
\Phi_{cd}(k;s_1, s_2)\,
\eeqra
and
\beq
 G_{ab} (k; \eta_a, \eta_b) = \left[g^{-1}_{ba} - 
\Sigma_{\vp_a \chi_b}\right]^{-1}(k; \eta_a, \eta_b)\,,
\label{fullG}
\eeq
where the last expression has to be interpreted in a formal sense, that is,
\[G_{ab} (k; \eta_a, \eta_b)=g_{ab} (\eta_a, \eta_b) + \int d s_1 d s_2 g_{ac} 
(\eta_a, s_1) 
\Sigma_{\vp_c \chi_d}(k; s_1, s_2) g_{db} (s_2, \eta_b) + \cdots\,.
\]
Since our goal is to compute the full non-linear power-spectrum, 
we see that our task reduces to that of computing the two functions 
$G_{ab}$ and $\Phi_{ab}$.

\section{Renormalization Group Equations}
The starting point of our formulation of the RG is a modification of 
the primordial power-spectrum appearing in
the path integral of Eq.~(\ref{genf}), as follows,
\beq
P^0(k) \rightarrow P^0_\l (k) = P^0(k) \,\Theta(\l,\,k)\,,
\label{Plam}
\eeq
where $ \Theta(\l,\,k)$ is a low-pass filtering function, which equals 
unity for $k \ll \l$ and zero for $k\gg \l$. In the computations
in this paper we will use a step function, 
\beq
\Theta(\l,\,k) = \theta(\l-k)\,,
\eeq
but more smooth behaviors, {\it e.g.} 
exponentials or power-laws, can be employed as well.

The modified generating functional, which we now indicate with  
$Z_\l[J_a,\, K_b;\,P^0] 
\equiv Z[J_a,\, K_b;\,P^0_\l]$, describes a fictitious Universe, 
in which the statistics of the initial conditions 
is modified by damping all fluctuations with momenta larger than $\l$. 
On the other hand, 
the dynamical content, encoded in the linear propagator and in the 
structure of the interaction, is left untouched. 
 
In the $\l \to \infty$ limit all the fluctuations are included, and
 we recover the physical situation. Increasing
the cutoff from $\l=0$ to $\l\to \infty$, the linear and non-linear 
effect of higher and higher fluctuations is gradually taken
into account. This process is described by a RG equation which can 
be derived by taking the $\l$ derivative
 of $Z_\l$,
\beqra
 \partial_\l Z_\l &=& -\frac{1}{2}  \int {\cal D}  \vp_a {\cal D} \chi_b
\;\exp{\{\cdots \}} \times\nonumber\\
&&\qquad\int d\eta_a\,d\eta_b d^3\bq\,\,\delta(\l-q) 
 P^0_{ab}(q) \delta(\eta_a) \delta(\eta_b)
 \chi_a(\eta_a) \chi_b(\eta_b),\nonumber\\
  &=&\frac{1}{2} \int d\eta_a\,d\eta_b d^3\bq\,\,\delta(\l-q) 
 P^0_{ab}(q) \delta(\eta_a) \delta(\eta_b)
 \frac{\delta^2 Z_\l }{\delta K_b \delta K_a} \,,
\label{ZRG}
\eeqra
where the meaning of the dots in the argument of the exponential in 
the first line can be read from Eqs.~(\ref{genf}) and (\ref{Plam}).

Starting from the equation above and the definition (\ref{WW}) 
one can obtain the RG equation for 
$W_\l$:
\beqra
 \partial_\l W_\l =&& \frac{1}{2} \int d\eta_a\,d\eta_b d^3\bq \,\delta(\l-q) 
 P^0_{ab}(q) \delta(\eta_a) \delta(\eta_b)
 \left(i \chi_b \chi_a + \frac{\delta^2 W_\l }{\delta K_b \delta K_a} 
\right)\,.\nonumber\\
 &&\nonumber\\
\label{WRG}
\eeqra

A similar equation can be obtained for the effective action 
$\Gamma_\l$. It is convenient to 
isolate the free from the interacting part, by writing
\beqra
\Gamma_\l [\vp, \,\chi]&&=  \int d\eta_a\,d\eta_b d^3\bq \left[
\frac{i}{2} \chi_a  P^0_{ab,\,\l}(q) \delta(\eta_a) \delta(\eta_b)\chi_b  
+ \chi_a g^{-1}_{ab} \vp_b 
\right] \nonumber\\
&&\quad+ \Gamma_{\mathrm{int},\l}[\vp, \,\chi]\,,\nonumber \\
&&\equiv  \Gamma_{\mathrm{free},\l}[\vp, \,\chi]+ 
\Gamma_{\mathrm{int},\l}[\vp, \,\chi]\,.
\eeqra
Using (\ref{WRG}), one obtains the RG equation for the interacting part 
of the effective action, 
which can be written in compact form as
\beq
 \partial_\l \Gamma_{\mathrm{int},\l} = 
 \frac{i}{2} \Tr \left[\left( \partial_\l
 {\bf \Gamma}^{(2)}_{\mathrm{free},\l}\right) \cdot \left( {\bf 
\Gamma}^{(2)}_{\mathrm{free},\l}+ {\bf \Gamma}^{(2)}_{\mathrm{int},\l} 
\right)^{-1}
  \right]\,,
  \label{gammaRG}
  \eeq
where ${\bf \Gamma}^{(2)}_{\mathrm{free},\l}$ is the matrix of 
second derivatives of the free action ({\it i.e.} 
the first terms on the LHS's of Eq.~(\ref{G2}), with $P^0\to P^0_\l$), 
while  ${\bf \Gamma}^{(2)}_{\mathrm{int},\l}$ is the matrix of 
second derivatives of the interacting action, 
computed at arbitrary field values. The trace indicates $\eta$ and 
momentum integrations, 
as well as summation over the doublet indices of Eq.~(\ref{doppietto}) 
and over the $\vp$ 
and $\chi$ field 
contributions. Notice that, since only 
the ``$\chi-\chi$'' entry of ${\bf \Gamma}^{(2)}_{\mathrm{free},\l}$ 
contains the 
theta function, through the primordial power-spectrum $P^0_\l$ of 
Eq.~(\ref{Plam}),
this is the only one contributing to the matrix 
$\partial_\l{\bf \Gamma}^{(2)}_{\mathrm{free},\l}$. 

In the next section, we will take a closer look at these equations by 
considering the specific example of the RG equation for
the full propagator $G_{ab,\l}$.

\section{The propagator and the emergence of an intrinsic UV cutoff}
\label{propo}
\subsection{Exact RG equation}
\begin{figure}
\centerline{\includegraphics[width = 4in , 
keepaspectratio=true]{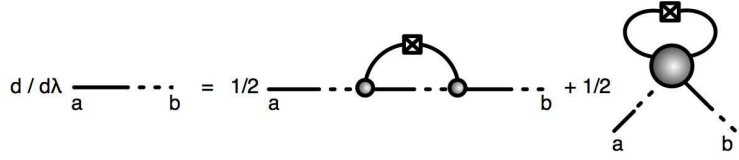}}
\caption{RG equation for the propagator $G_{ab,\l}$}
\label{prop_RG}
\end{figure}

The RG equation for the propagator defined in Eq.~(\ref{w2}), is 
derived by taking the appropriate double derivative of. eq~(\ref{WRG}),
\beqra
&& \partial_\l\left. \frac{\delta^2 W_\l}{\delta J_a(\bk,\,\eta_a)
 \delta K_b(\bk^\prime,\,\eta_b)} \right|_{J_a,\,K_b=0}= 
-\delta(\bk+\bk^\prime) \,\partial_\l G_{ab\,,\l} (k,\,\eta_a,\,\eta_b) = 
\nonumber\\ 
 &&\nonumber\\
&& \frac{1}{2} \int d\eta_c\,
 d\eta_d \,d^3\bq \,\delta(\l-q) 
   P^0_{cd}(q) \delta(\eta_c) \delta(\eta_d)
\left. \frac{\delta^4 W_\l }{\delta J_a \,\delta K_d 
\,\delta K_c\,\delta K_b}\right|_{J_a,\,K_b=0}\,,
\label{RGprop}
\eeqra
were we have used the fact that $ \chi_a$ and
 $\delta \chi_a/\delta K_b = \delta^2 W/ \delta K_a \delta K_b$ 
both vanish for vanishing sources $J_a$ 
 and $K_b$.
 The full connected four-point function, 
  \beq
\delta(\bk_a+\bk_b+\bk_c+\bk_d)\,  W^{(4)}_{J_a K_b K_c K_d,\l} 
\equiv  \left. \frac{\delta^4 W_\l }{\delta J_a 
\,\delta K_d \,\delta K_c\,\delta K_b}\right|_{J_a,\,K_b=0}\,,
  \eeq
 can be written in terms of
  full 1PI three and four-point functions and full propagators, as
 \beqra
 \label{W4}
&& W^{(4)}_{J_a K_b K_c K_d,\l}(\bk,\eta_a;-\bk,\eta_b;\bq,\eta_c;-\bq,\eta_d) 
\nonumber\\
&&=\int ds_1\cdots ds_4 \,  G_{ae,\l}(k;\eta_a,s_1) 
G_{fb,\l}(k;s_2,\eta_b)G_{gc,\l}(q;s_3,\eta_c)G_{hd,\l}(q;s_4,\eta_d)  
\nonumber \\
&&  \times\biggl\{\Gamma^{(4)}_{\chi_e \vp_f \vp_g 
\vp_h,\l}(\bk,s_1;-\bk,s_2;\bq,s_3;-\bq,s_4)  \nonumber\\
&&- \int ds_5 ds_6 \; G_{li,\l}(k-q;s_5,s_6)\nonumber
\\ && 2\, \Gamma^{(3)}_{\chi_e\vp_h\vp_l,\l}(\bk,s_1;-\bq,s_4;-\bk+\bq,s_5)
\Gamma^{(3)}_{\chi_i\vp_g\vp_f,\l}(\bk-\bq,s_6;\bq,s_3;-\bk,s_2)
\biggr\},
  \eeqra
 where we have defined
 \beq
\delta(\bk_a+\bk_{b_1}+\cdots+\bk_{b_{n-1}}) 
\,\Gamma^{(n)}_{\chi_a \vp_{b_1}\cdots \vp_{b_{n-1}},\l} 
\equiv \left.\frac{\delta^{n}\Gamma_\l}{\delta \chi_a  \delta
 \vp_{b_1}\cdots \delta \vp_{b_{n-1}}}\right|_{\chi_a,\vp_b=0}\,.
\eeq
Inserting the expression above in Eq.~(\ref{RGprop}) we obtain the RG
equation represented in Fig.~\ref{prop_RG}, where the thick lines indicate full
propagators, 
dark blobs are full 1PI 3 and 4-point functions, and the 
crossed box is the {\em RG kernel }
\beq
\label{kernel}
K_{gh,\l}(q;s_3,s_4) = G_{gc, \l}(q;s_3,0)
G_{hd, \l}(q;s_4,0) P^0_{cd}(q)\,\delta(\l-q) \;.
\eeq
Notice that the kernel can be obtained by deriving w.r.t. $\l$ the $\theta$ 
function multiplying the primordial power-spectrum in $P^I_\l$, 
see Eqs.~(\ref{fullP1}) and (\ref{Plam}). 

The RHS of the RG equation is remarkably simple. The two contributions 
have just the structure of 1-loop diagrams, where the tree-level 
vertices and propagators have been replaced by full, $\l$-dependent ones. 
The same holds true not only in the case of the propagator, but for any 
other quantity one is interested in.

A recipe can be given to obtain the RG equation for any given 
quantity: 

\begin{itemize}
\item write down the {\em 1-loop expression} 
for the quantity of interest, obtained using  {\em any needed vertex},
(for instance, in Fig.~\ref{prop_RG},
we have not only the vertex $\chi \vp \vp$, but 
also 
$\chi \vp\vp\vp$, although it vanishes at tree-level);  
\item promote the linear propagator, the power-spectrum and the 
vertices appearing in 
that expression to full, $\l$-dependent ones;
\item take the $\l$-derivative of the full expression, 
by considering only 
the explicit $\l$-dependence of the step-function contained in 
$P^I_\l$.
\end{itemize} 
It should be emphasized that the RG equations obtained following 
these rules are {\em exact}, in the sense that they encode all the 
dynamical and statistical content of the path-integral (\ref{genf}) 
or, equivalently, of the continuity, Euler and Poisson equations 
supplemented by the initial power-spectrum. Despite their 1-loop structure, 
the complete RG equations resum the theory at
any loop order, since the ($\lambda$-dependent) full propagators 
and n-point functions appearing at the RHS contain the 
contributions of all orders of perturbation theory \cite{RGreview}.
\subsection{Approximations}
Different approximation schemes can be attempted. Although the RG
equations can be solved perturbatively, thus reproducing the results 
of PT, they are most indicated for non-perturbative 
resummations. 
As a first step, we note that the full 3 and 4-point functions 
appearing in the RG equation for $G_{ab,\l}$  are
also $\l$-dependent quantities, which can be computed by RG equations, 
also derived from Eqs.~(\ref{WRG}, \ref{gammaRG}). These equations depends, 
in turn, 
on full, $\l$-dependent, functions up to 5 
(for the 3-point function) or 6 
(for 4-point ones) external legs, which also evolve according 
to RG equations.  Approximations to the full RG flow then amount
to truncating the full hierarchy of coupled differential equations, and 
using some ansatz for the full n-point functions appearing in the
surviving equations. 

In this paper, we will approximate the full RG flow by keeping
the running of the two 2-point functions (propagator and power-spectrum) and
keeping the tree-level expression for the trilinear vertex $\chi\vp\vp$, 
{\it i.e},
\beqra
\label{trezero}
&&\Gamma^{(3)}_{\chi_a\vp_b\vp_c,\l}(\bk,s_1;-\bq,s_2;-\bk+\bq,s_3)\nonumber\\
&& \quad\qquad\simeq -2\, \delta(s1-s) \delta(s_2-s)\delta(s_3-s)\, e^s\,
 \gamma_{abc}(\bk,-\bq,-\bk+\bq)\,,
 \eeqra
 while all the other $n$-point functions appearing at the RHS of the RG 
 equations are kept to zero, since they vanish at tree-level.
In this approximation, only the first diagram on the RHS of Fig.~\ref{prop_RG}
contributes to the running.

\subsection{1-loop}
Since the RG equations have the structure of 1-loop integrals, PT can be reproduced by
solving them iteratively.  The $l$-th order contribution to a
given quantity can be obtained by using up to $m$-th order functions 
(with $m\le l-1$) in 
the RHS of the
  corresponding RG
  equation and then closing the loop with the kernel 
  (\ref{kernel}), evaluated at order $l-m-1$. 
 So, the computation of the 1-loop contribution to the propagator  
requires the tree-level
 expression for $W^{(4)}_{J_a K_b K_c K_d,\l}$, obtained by setting all 
the quantities appearing in Eq.~(\ref{W4}) at their tree-level values, that is
\beqra
&&  W^{(4),\mathrm{tree}}_{J_a K_b K_c K_d,\l} =   -8 \int d s_1 d s_2  
\,e^{s_1+ s_2} \,g_{hd}(s_1-\eta_d) \,
g_{gc}(s_2-\eta_c) g_{ae}(\eta_a-s_1)\,    \nonumber \\ 
&&   \gamma_{ehl}(\bk, -\bq, -\bk + \bq) \,g_{li}(s_1-s_2)\,
\gamma_{igf}(\bk-\bq, \bq, -\bk) \,g_{fb}(s_2-\eta_b).
\label{tree4}
\eeqra


The above expression is $\l$-independent, so that, once inserted in 
(\ref{RGprop}), it 
allows a straightforward integration in $\l$  from $\l=0$ to $\l=\infty$. 
The initial condition at $\l=0$ is 
\beq
G_{ab,\,\l=0} ( k;\,\eta_a,\,\eta_b)= g_{ab} (\eta_a,\,\eta_b)\,,
\label{initialG}
\eeq
and the well-known 1-loop result, discussed for instance in 
Ref.~\cite{Crocce2}, 
is reproduced\footnote{Since
our definition of the tree level propagator differs by that of 
Ref.~\cite{Crocce2} by a $e^{-\eta}$ 
factor, see Eqs.~(\ref{doppietto}, \ref{compact}), our result for the 
1-loop propagator has to be multiplied
by $e^{(\eta_a-\eta_b)}$ before comparing with \cite{Crocce2}.}.
\subsection{Large $k$ resummation} 
A better approximation to the full RG solution is obtained by keeping
at least some of the $\l$-dependence on the RHS of Eq.~(\ref{RGprop}). 
An analytic result can be
obtained in the $k \gg \l$ limit. In this regime, we can still approximate
 the kernel (\ref{kernel}) -- which carries
momentum $q=\l$ -- with its linear expression, 
\beq
K_{gh,\l}(q;s_3,s_4) \simeq u_g u_h\, \theta(s_3)\theta(s_4)\,P^0(q)\,
\delta(\l-q) \;,
 \label{p1}
\eeq
where we have put the initial conditions on the growing mode, {\it i.e.}, 
$w_a=u_a$, with $u_a$ given in Eq.~(\ref{ic}). 

Starting from Eq.~(\ref{vertice}) one can verify that, 
in the $k\gg q=\l$ limit 
\cite{Crocce2}, 
\beq
u_f\,\gamma_{efg}(-\bk,\,\bq,\,\bk-\bq) \simeq \delta_{eg} \,
\frac{1}{2}\frac{k}{q}\, \cos \bk\cdot\bq\,,
\label{vertk}
\eeq
so that Eq.~(\ref{tree4}), once inserted in (\ref{RGprop}), gives
\beqra
 &&\partial_\l G_{ab\,,\l} (k;\,\eta_a,\,\eta_b) = - g_{ab}(\eta_a,\,\eta_b) 
\,k^2 
 \,\nonumber\\
&& \qquad\qquad \times \int_{\eta_b}^{\eta_a} ds_2 \,
\int_{\eta_b}^{s_2} ds_1 \,e^{s_1+s_2}\,\int d^3\bq\, \delta(\l-q) P^0(q)   
\frac{(\cos \bk\cdot\bq)^2}{q^2}  \nonumber\\ 
&&  \qquad\qquad = - g_{ab}(\eta_a,\,\eta_b) \, \frac{k^2}{3} \,
 \frac{\left(e^{\eta_a} -e^{\eta_b}\right)^2}{2}\int d^3\bq
\,\delta(\l-q) \,\frac{P^0(q)}{q^2}\,,
\eeqra
where we have used the following property of the propagators,
\beq
 g_{ae}(\eta_a,\,s_2) g_{el}(s_2,\,s_1)= g_{al}(\eta_a,\,s_1)\, 
\theta(\eta_a-s_2) \theta(s_2-s_1) .
 \label{propprop}
 \eeq
The equation above is just the large $k$ limit of the 1-loop 
result discussed in the previous subsection. 
Now, a first level of RG improvement  consists in promoting
the linear propagator $g_{ab}(\eta_a,\eta_b)$ on the RHS 
above to the full --
$k$ and $\l$-dependent-- one $G_{ab\,,\l} (k;\,\eta_a,\,\eta_b)$, 
while keeping the tree-level expression for the kernel and for the vertices. 
In this approximation, the RG equation can be integrated analytically, 
giving
\beq
G_{ab\,,\l} (k;\,\eta_a,\,\eta_b) =g_{ab}(\eta_a,\,\eta_b) 
e^{ - \frac{k^2}{3} \,
 \frac{\left(e^{\eta_a} -e^{\eta_b}\right)^2}{2}\int d^3\bq
\,\theta(\l-q) \,\frac{P^0(q)}{q^2} }\,,
\label{RGres}
\eeq
where the same initial condition at $\l=0$, Eq.~(\ref{initialG}), 
has been imposed.

\begin{figure}
\centerline{\includegraphics[width = 6in , 
keepaspectratio=true]{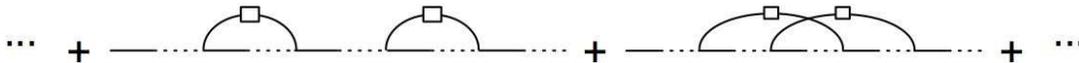}}
\caption{The infinite class of diagrams resummed by the RG in the approximation leading to Eq.~(\ref{resu}). }
\label{prinpath}
\end{figure}

In the $\l\to \infty$ limit, the above expression gives
\beq G_{ab\,,\l} (k;\,\eta_a,\,\eta_b) =g_{ab}(\eta_a,\,\eta_b) 
e^{ - k^2 \sigma_v^2 \,
 \frac{\left(e^{\eta_a} -e^{\eta_b}\right)^2}{2} }\,,
 \label{resu}
\eeq
where $\sigma_v^2$ is the velocity dispersion, defined as
\beq
\sigma_v^2 \equiv \frac{1}{3} \int d^3\bq
\, \,\frac{P^0(q)}{q^2}\,.
\eeq
Two comments are in order at this point.
First, the RG improvement discussed here has a clear interpretation in
terms of PT. Indeed, 
as shown in \cite{Crocce2}, the result (\ref{resu}) can be obtained
also  by resumming the infinite set of diagrams 
shown in Fig.~\ref{prinpath}, in which all the power-spectra are 
directly connected to the propagator line carrying momentum $k$. 
It is amazing how the same result, that in PT requires
a careful control of the combinatorics, is here obtained by a simple, 
1-loop, integration.

The second, more physical,
comment has to do with the dramatic modification of the UV
behavior of the resummed propagator w.r.t. 
the linear one, and on its impact on the RG flow. 
Indeed, when the propagator of Eq.~(\ref{resu}) is employed in the full kernel 
$K_{gh,\l}$ of Eq.(\ref{kernel}), an intrinsic UV cutoff is provided 
to the RG flow: fluctuations 
with large momenta are exponentially damped, so that the RG 
evolution freezes out for $\l \gg e^{-\eta} /\sigma_v$. 
This is a genuinely non-perturbative effect, 
which is masked if one considers PT at any finite
 order. Indeed, expanding the exponential 
in Eq.~(\ref{resu}), one can see that higher 
orders in PT are more and more divergent in the UV, 
obtaining just the opposite behavior.
In other words, the full theory is much better behaved 
in the UV than any of its perturbative approximations. 

\subsection{RG result}
\begin{figure}
\centerline{\includegraphics[width = 4in , 
keepaspectratio=true]{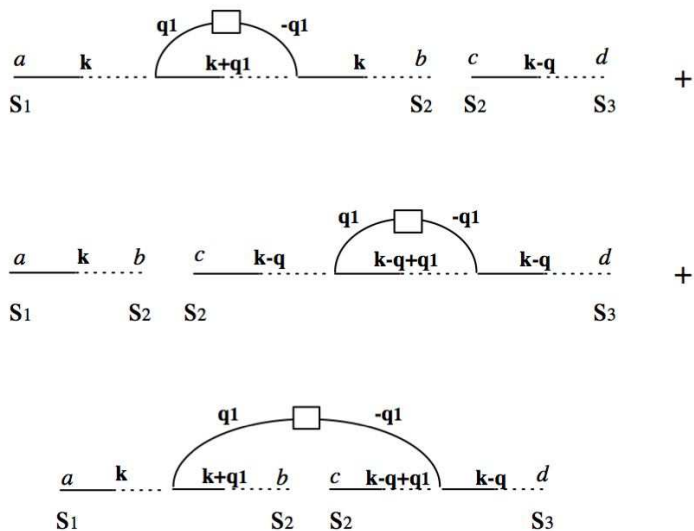}}
\caption{The 1-loop contribution to the combination of Eq.~(\ref{combination}). }
\label{1loopcomb}
\end{figure}

An improvement on the approximations leading to Eq.~(\ref{resu}) consists 
in relaxing the $k\gg \l$ condition and in keeping
some amount of $\l$-dependence also in the propagators appearing in the 
kernel. 
In this section, we consider the RG equation obtained from Eq.~(\ref{RGprop})
by  keeping full, $\l$-dependent, propagators in Eq.~(\ref{W4}), while
leaving the vertex functions at their tree-level expression, 
Eq.~(\ref{trezero}), with their full-momentum dependence taken 
into account, that is, not using the approximation in Eq.~(\ref{vertk}). 
Consistently, we will also take $\Gamma^{(4)}_{\chi\vp\vp\vp,\l}=0$. 
Higher order approximations, 
in particular including
the running of the vertex, will be considered elsewhere.
We have to deal with combinations of full propagators 
and tree-level vertices as
\beq
G_{ab,\l}(k;\,s_1,\,s_2) \gamma_{bfc}(-\bk,\,\bq,\,\bk-\bq) 
G_{cd,\l}(|\bk-\bq|;\,s_2,\,s_3)\,.
\label{combination}
\eeq
At tree-level, it can be written as
\beqra
&&
g_{ab}(s_1,\,s_2) \gamma_{bfc}(-\bk,\,\bq,\,\bk-\bq) 
g_{cd}(s_2,\,s_3) \equiv \nonumber \\
&& B_{afc}(\bk,\,\bq)g_{cd}
(s_1,\,s_3)\, \theta(s_1-s_2) \theta(s_2-s_3) \,,
\label{combtree}
\eeqra
where the functions $B_{afc}$ keep track of the full momentum and 
matrix structure of the tree-level trilinear vertex. The 1-loop 
contribution to the above combination is given by the sum of the 
three diagrams in Fig.~\ref{1loopcomb}, which in the $\l \ll k$ limit gives
\beqra 
&&-  \frac{k^2}{3} \,
 \frac{\left(e^{s_1} -e^{s_3}\right)^2}{2}\int d^3\bq_1
\,\theta(\l-q_1) \,\frac{P^0(q_1)}{q_1^2} \times \nonumber\\
&& \qquad\qquad\qquad\qquad 
g_{ab}(s_1,\,s_2) \gamma_{bfc}(-\bk,\,\bq,\,\bk-\bq) 
g_{cd}(s_2,\,s_3)\nonumber\\
&&=
-  \frac{k^2}{3} \,
 \frac{\left(e^{s_1} -e^{s_3}\right)^2}{2}\int d^3\bq_1
\,\theta(\l-q_1) \,\frac{P^0(q_1)}{q_1^2} \times \nonumber\\
&& \qquad\qquad\qquad\qquad B_{afc}(\bk,\,\bq)g_{cd}
(s_1\,s_3)\, \theta(s_1-s_2) \theta(s_2-s_3)\,.
\eeqra
Notice that the third diagram in Fig.~\ref{1loopcomb} actually amounts to a 1-loop correction to the trilinear vertex. Repeating the procedure
at $n$-loops we realize that the corrections can be resummed at any order, giving 
\beq
 B_{afc}(\bk,\,\bq)g_{cd}
(s_1,\,s_3)\,e^{ - \frac{k^2}{3} \,
 \frac{\left(e^{s_1} -e^{s_3}\right)^2}{2}\int d^3\bq
\,\theta(\l-q) \,\frac{P^0(q)}{q^2} } \theta(s_1-s_2) \theta(s_2-s_3)\,,
\eeq
where we recognize the resummed propagator of Eq.~(\ref{RGres}). Therefore, we learn that the resummation of loop corrections to the combination (\ref{combtree}) {\em is not} the product of two resummed propagators, but a single resummed propagator multiplied by the $B_{afbc}$ matrix.  In particular, the dependence on the intermediate `time' $s_2$ appears only in the causal structure enforced by the $\theta$ functions, while it is absent from the exponential decay factor.
We  will assume this property to hold also for the RG-resummed propagator, that is, at the RHS of RG equations we will make use of the relation  
\beqra
&& G_{ab,\l}(k;\,s_1,\,s_2) \gamma_{bfc}(-\bk,\,\bq,\,\bk-\bq) 
G_{cd,\l}(|\bk-\bq|;\,s_2,\,s_3)
=  \nonumber \\
&& B_{afc}(\bk,\,\bq)G_{cd,\l}(k;\,s_1,\,s_3)\, \theta(s_1-s_2) \theta(s_2-s_3) 
\,.
\label{compo}
\eeqra
Our purpose here will be to go beyond the large $k$ resummation of Eq.~(\ref{resu}). Therefore we will allow an extra $k$ dependence of the propagator, to be determined by the RG equations. We will therefore use the following ansatz for the full propagator
\beq
G_{ab,\l}(k;\,s_1,\,s_2) = H_{ab,\l}(k) \,e^{ - k^2 \sigma_v^2 \,
 \frac{\left(e^{s_1} -e^{s_2}\right)^2}{2} }\,,
 \label{ansatzG}
 \eeq
in which we have factored out the leading time behavior of Eq.~(\ref{resu}), 
and we will use the RG to compute deviations of $H_{ab,\l}(k)$ from unity.
Inserting the expression above in the RG equations one obtains, 
for each external momentum $k$, a closed system of four coupled 
differential equations, one for 
each component of $H_{ab,\,\l}$. In order to simplify it further, 
we will consider the two combinations $G_{ab,\l}u_b=G_{a1,\l}+G_{a2,\l}$,
($a=1,\,2$), therefore we will solve the equations for
\beq
h_{a,\l}(k)= H_{ab,\l}(k)\, u_b\,,\qquad\qquad(a=1,\,2)\,.
\label{hsmall}
\eeq
These are given by
\beq
 \partial_\l h_{a,\l}(k) = - 4\pi k^2 \, 
 S[\l;\,\eta_a,\,\eta_b] \,M_a\left[\l/k\right] \, P^0(\l) 
\,\bar{h}_\l^2(\l) \, \bar{h}_\l(k)\,,
 \label{hRG}
 \eeq
where we have approximated all the $h_{a,\,\l}$'s appearing at 
the RHS with
\beq
h_{a,\l}(q) \simeq  u_a \bar{h}_\l(q) \equiv u_a\,\frac{h_{1,\l}(q) 
+h_{2,\l}(q) }{2}\,,
\label{rightapprox}
\eeq
and we have defined
\beqra
 && M_1[r] = \nonumber\\
&&-\frac{1}{252 \,r^3}\left[r(6-79r^2+50r^4-21r^6) 
+\frac{3}{2}(1-r^2)^3 (2+7r^2) \log\left|\frac{1-r}{1+r}
\right| \right]\,,\nonumber\\
&&\nonumber\\
 && M_2[r] = \nonumber\\
&&-\frac{1}{84 \,r^3}\left[ r(6-41r^2+2 r^4- 3 r^6) 
+\frac{3}{2}(1-r^2)^3 (2+ r^2) \log\left|\frac{1-r}{1+r}
\right| \right]\,.
\eeqra
The `time' dependence is contained in the function
\beqra 
S[\l;\eta_a,\,\eta_b] &=&\int_{\eta_b}^{\eta_a}ds_2
\int_{\eta_b}^{s_2}ds_1\,e^{s_1+s_2} 
\,e^{-\frac{\l^2\sigma_v^2}{2}[(e^{s_1}-1)^2+(e^{s_2}-1)^2]}\nonumber\\
 &=&\frac{\pi}{4 \l^2 \sigma_v^2} \left\{{\mathrm{erf}}
\left[ \frac{\l\,\sigma_v (e^{\eta_a}-1)}{\sqrt{2}}\right]-
{\mathrm{erf}}\left[ \frac{\l\,\sigma_v (e^{\eta_b}-1)}{\sqrt{2}}\right]
\right\}^2\,,
\eeqra
with $\mathrm{erf}$ the `error function', $\mathrm{erf}(z)=2 \pi^{-1/2} 
\int_0^z \,dt \,e^{-t^2}$. 
Notice that in the limit, $\sigma_v\to 0$, we recover the 1-loop
behavior, $ S[\l;\eta_a,\,\eta_b] \to (e^{\eta_a} -e^{\eta_b})^2/2$.

\begin{figure}
\centerline{\includegraphics[width = 4in , 
keepaspectratio=true]{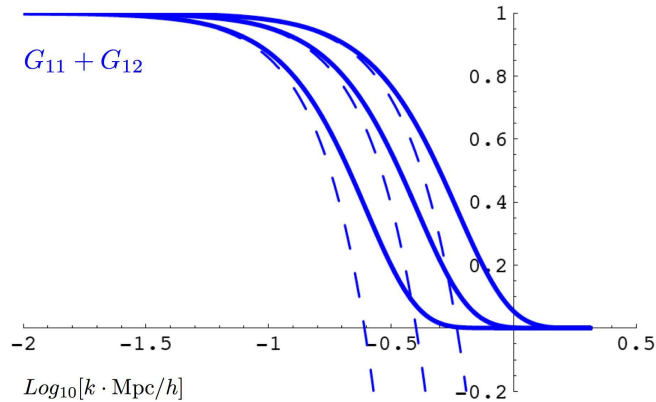}}
\caption{The propagator in a $\Lambda$CDM model at 
$z=0,1,2$, from left to right. Solid lines represent the results
of the integration of Eq.~(\ref{hRG}). Dashed lines are 1-loop results. }
\label{resProp}
\end{figure}

Notice that, due to the difference between $M_1[r]$ and
$M_2[r]$, we have two different equations for $h_{1,\l}$ and 
$h_{2,\l}$, {\it i.e.} two different propagators for the density and the 
velocity divergence, even if we have approximated them according 
to Eq.~(\ref{rightapprox}) on the RHS.

The initial conditions of the RG at $\l=0$ can be read from 
Eq.~(\ref{initialG}), namely
\beq
h_{a,\l=0}(k) = u_a\, e^{ k^2 \sigma_v^2 \,
 \frac{\left(e^{\eta_a} -e^{\eta_b}\right)^2}{2}}\,.
 \eeq
 
The RG equations  (\ref{hRG}), in the limit $\sigma_v\to 0$ and $\bar{h}_\l 
\to1$, give the
1-loop expression for the propagators.

We consider a spatially flat $\L$CDM model 
with $\Omega_\L^0=0.7$, $\Omega_b^0=0.046$, $h=0.72$, $n_s=1$. 
The primordial power-spectrum $P^0$ is taken from the output of linear 
theory at 
$z_{in}=80$, as given by the CAMB Boltzmann code \cite{CAMB}.

In Fig. \ref{resProp} we plot our results for the propagator. 
Linear theory, Eq.~(\ref{proplin}), corresponds to $G_{11}+G_{12}=1$ 
for any external momentum $k$. The solid lines represent the results
of the integration of the RG equation (\ref{hRG}) at redshifts 
$z=0,1,2$, from left to right. We plot also the 1-loop results, given
by the dashed lines. Notice that, as $k$ grows, the latter become
negative, signaling the breakdown of the perturbative expansion.

As one can see, the damping of the propagator for large $k$, that we obtained
in the large momentum approximation leading to Eq.~(\ref{resu}),
is exhibited also by the solutions of the more accurate RG equations
(\ref{hRG}).

\section{The power-spectrum}
The full power-spectrum has the structure of 
Eqs.~(\ref{fullP1}, \ref{fullP2}), where the primordial power-spectrum 
$P^0$ contained 
 in $P^I_{ab}$ has been 
multiplied by $\theta(\l-k)$, as in Eq.~(\ref{Plam}), and
the full propagators and the function $\Phi_{ab}$ are now 
$\l$-dependent quantities.
The RG evolution of the first term, $P^I_{ab,\l}$, is completely 
determined by that of the full propagator, that
we have computed in the previous section. 
\begin{figure}
\centerline{\includegraphics[width = 4in 
, keepaspectratio=true]{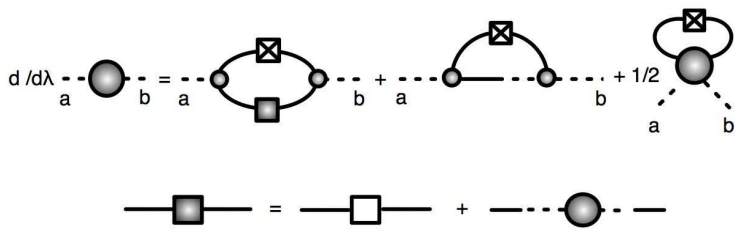}}
\caption{RG equation for $\Phi_{ab,\l}$}
\label{runPhi}
\end{figure}
Our task now reduces to computing the evolution
of $P^{II}_{ab,\l}$, 
\beqra
\partial_\l &&P^{II}_{ab,\l}(k; \eta_a, \eta_b)= \nonumber\\
 &&\int_0^{\eta_a} ds_1
\int_0^{\eta_b} ds_2 \biggl[
\partial_\l  G_{ac,\l}(k;\eta_a,s_1)\,
G_{bd,\l}(k;\eta_b,s_2) 
\Phi_{cd,\l}(k;s_1,s_2) \nonumber \\
& &\qquad\qquad\qquad+ G_{ac,\l}(k;\eta_a,s_1)\,
\partial_\l G_{bd,\l}(k;\eta_b,s_2) 
\Phi_{cd,\l}(k;s_1,s_2) \nonumber \\
& &\qquad\qquad\qquad+ G_{ac,\l}(k;\eta_a,s_1)\,
G_{bd,\l}(k;\eta_b,s_2) 
\partial_\l  \Phi_{cd,\l}(k;s_1,s_2)\biggr]\,.
 \label{RGP2}
\eeqra 
The missing element is the running the 1PI two-point function, 
$ \Phi_{ab,\,\l}$, which can be obtained by the recipe given in 
sect.~\ref{propo} and is represented graphically in Fig.~\ref{runPhi}.
The dark box represents the full, $\l$-dependent, power-spectrum.
Notice that, besides the $\chi\vp\vp$ vertex, also $\chi\chi\vp$ and 
$\chi\chi\vp\vp$ appear in the full RG equation.

We will stick to the same truncation scheme considered in the computation of
the propagator, that is, we will allow running propagator and 
power-spectrum and will keep the vertices at tree-level. In this approximation,
only the first diagram at the RHS of Fig.~\ref{runPhi} contributes to 
the running,
and the RG equation is
\beqra
&&  \partial_\l  \Phi_{ab,\,\l}(k;\,s_1,\,s_2) = 4 \,e^{s_1+s_2} \, 
\int d^3 \bq \,\delta(\l-q)
P^I_{dc,\l}(q; \,s_1, \,s_2)\times\nonumber\\
&&
\qquad   P_{fe,\l}(|\bq-\bk|; \,s_1, \,s_2)\,
 \gamma_{adf}(\bk,-\bq,-\bk+\bq) \gamma_{bce}(-\bk,\bq,\bk-\bq) \,.
 \label{RGPhi}
  \eeqra 
We will proceed along the same lines
we followed in the computation of the propagator. 
Namely, we will again make use of the 
  ansatz (\ref{ansatzG})
for the propagator, and of the approximation (\ref{rightapprox}) 
for those propagators appearing on the RHS of the RG equations. 
As a consequence, the contribution $P^I_{ab,\l}$ to the power-spectrum 
will be approximated, at the RHS, as
\beq
 P^I_{ab,\l}(q; \,\eta_a, \,\eta_b) \simeq u_a u_b\, \bar{h}_\l^2(q) P^0(q) 
 e^{ - q^2 \sigma_v^2 \,
 \frac{\left(e^{\eta_a} -1\right)^2+\left(e^{\eta_b}-1\right)^2}{2}}\,.
 \label{P11}
\eeq
To reduce the number of equations, we focus on the evolution of 
an `average' $P^{II}_{ab,\l}$, defined as
\beq
\bar{P}^{II}_{\l}\equiv \frac{1}{4}\, u_a\,u_b\,P^{II}_{ab,\l}\,.
\eeq
Since $P^{II}_{ab,\l}$ evaluated at $s_1$ and $s_2$ appears on the RHS of 
Eq.~(\ref{RGPhi}) we need it at generic time arguments. 
While, in principle, the complete time dependence could be obtained 
from Eq.~(\ref{RGP2}) itself, it would be in practice extremely 
time-consuming. Therefore, we will make two different {\it ansatze} 
on the time dependence of the  $P^{II}_{ab,\l}$ appearing on the RHS of 
(\ref{RGPhi}), verifying {\it a posteriori} the stability of our results 
with respect to the two different choices.
The first ansatz is inspired by the analogy with Eq.~(\ref{P11}),
\beq
\bar{P}^{II}_{\l}(q; \,\eta_a, \,\eta_b) = 
\bar{P}^{II}_{\l}(q; \,s_1, \,s_2)\,e^{ - q^2 \sigma_v^2 \,
 \frac{\left(e^{\eta_a} -e^{s_1}\right)^2+
\left(e^{\eta_b}-e^{s_2}\right)^2}{2}}\,,
\label{ans1}
\eeq
while the second one is simply
\beq
\bar{P}^{II}_{\l}(q; \,\eta_a, \,\eta_b) = 
\bar{P}^{II}_{\l}(q; \,s_1, \,s_2)\,.
\label{ans2}
\eeq
The comparison between the power-spectra obtained with these two
choices will be given in Fig.~\ref{P2ansatz} below.

Moreover, we will compute the power-spectrum at equal times, 
that is $\eta_a=\eta_b=\eta$.

Inserting the expressions above in (\ref{RGPhi}), and then 
contracting Eq.~(\ref{RGP2}) with 
$u_a\,u_b$, we arrive, after performing the integrations in 
$s_1$, $s_2$, and $\bq$, to
\beqra
&&\partial_\l \bar{P}^{II}_{\l}(k;\eta)=
- 4\pi k^2 \,\bar{h}_\l^2(\l)  \, P^0(\l)\biggl\{
 S[\l;\,\eta,\,0] \,u_a\,M_a\left[\l/k\right] \,\, \bar{P}^{II}_{\l}(k;\eta) 
 \nonumber\\
 &&\qquad\qquad\quad- \frac{1}{2k} \int_{|k-\l|}^{k+\l} 
dp\biggl[\theta(\l - p) \bar{h}_\l^2(p) U(\l^2,k^2,p^2;\eta)
  P^0(p)  \nonumber\\
  && 
+U(k^2,\l^2,-p^2;\eta)  \bar{P}^{II}_{\l}(p;\eta)
 \biggr]  \frac{\left(2k^4-3(p^2-\l^2)^2+k^2(p^2+\l^2) 
\right)^2}{100 k^2p^3\l^3}
\biggr\}\,,
\label{P2RG}
\eeqra 
with the initial condition 
\beq
\bar{P}^{II}_{\l=0}(k;\eta) = 0\,.
\eeq
The function $U$ is defined as
\beqra
&& U(\l^2,k^2,p;^2\eta)= 
\biggl[
\int_0^\eta ds\, e^s \, e^{-\frac{(\l^2+p^2)\sigma_v^2}{2}(e^s-1)^2 
-\frac{k^2 \sigma_v^2 }{2}(e^\eta-e^s)^2}
\biggr]^2\nonumber\\ 
&&
\qquad\qquad\qquad= \frac{\pi}{2 \sigma_v^2} \frac{1}{\l^2+k^2+p^2}\,
e^{- \frac{k^2(\l^2+p^2)}{\l^2+k^2+p^2}\sigma_v^2(e^\eta -1)^2 }\times
 \nonumber\\
&&
 \left\{\mathrm{erf}\left[  \frac{k^2 \sigma_v}{\sqrt{\l^2+k^2+p^2}} 
\frac{(e^\eta-1)}{ \sqrt{2}} \right] 
 + \mathrm{erf}\left[  \frac{(\l^2+p^2)\sigma_v}{\sqrt{\l^2+k^2+p^2}} 
\frac{(e^\eta-1)}{ \sqrt{2}} \right] 
 \right\}^2\,,
\eeqra
and, in the $\sigma_v\to0$ limit it goes to  $(e^{\eta}-1)^2$. In this 
limit, and setting 
$\bar{P}^{II}_{\l}=0$ in the RHS, the integration of Eq.~(\ref{P2RG}) 
gives the 1-loop result.
\begin{figure}
\centerline{\includegraphics[width = 6.5in,
keepaspectratio=true
]{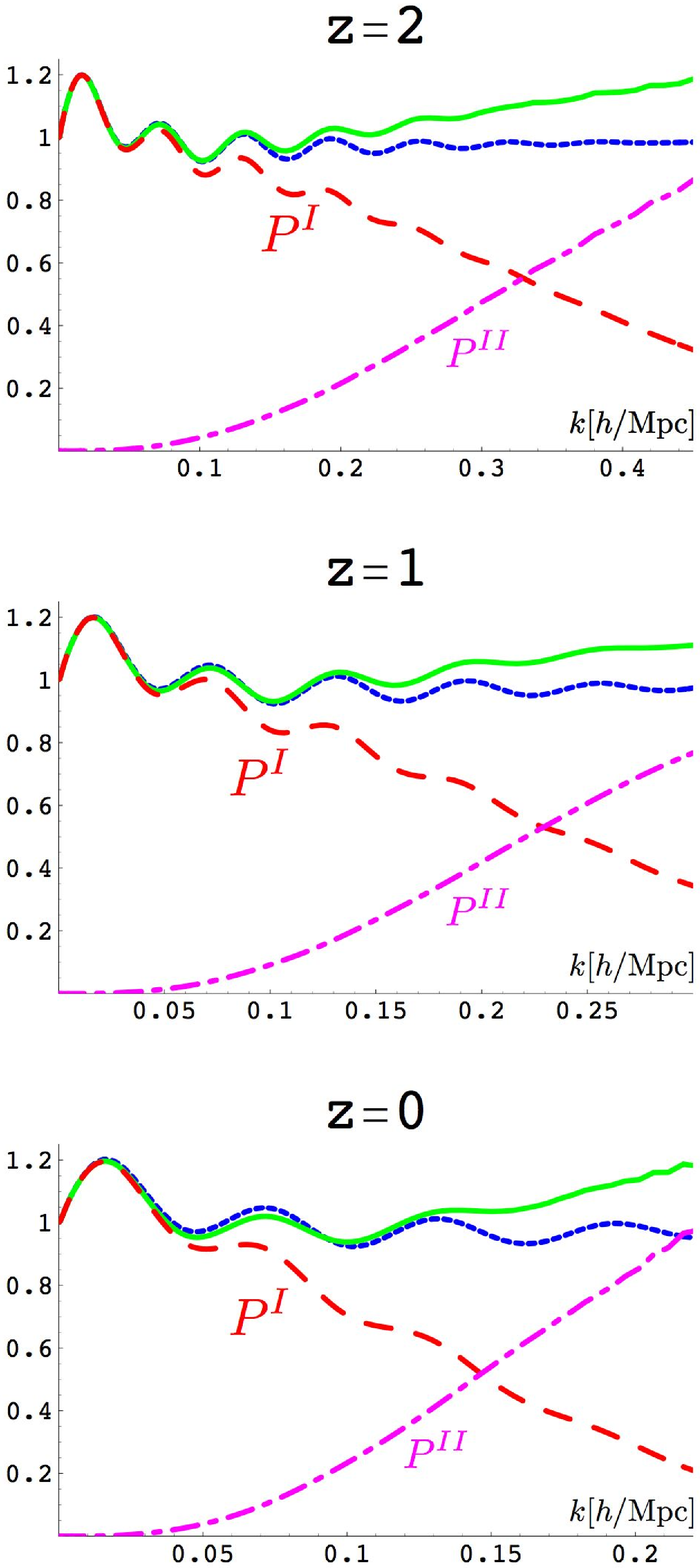}}
\caption{The contributions of $P^I_{11}$ (long-dashed line) 
and $P^{II}_{11}$ (dot-dashed) to the total 
power-spectrum, $P_{11}=P^I_{11}+P^{II}_{11}$ (solid). The 
linear theory result (short-dashed) is also shown.}
\label{P1P2}
\end{figure}

In Fig.~\ref{P1P2} we plot our results for the power-spectrum (solid lines), 
in the momentum range relevant for the BAO, at $z=0,\,1$, and $2$, 
from bottom to top. 
Long-dashed lines and dot-dashed ones correspond to the two 
contributions $P^I_{11}$ and $P^{II}_{11}$ to the full power-spectrum 
[see Eqs.~(\ref{fullP1}, \ref{fullP2})], respectively. The results of 
linear theory are given by the short-dashed lines. Notice the decrease 
of the contribution $P^I_{11}$ for growing $k$, due to the damping of the 
propagators seen in Fig.~\ref{resProp}. For the lower peaks, this is 
partially compensated by the contribution $P^{II}_{11}$, so that the full
power-spectrum approximately tracks the linear theory result up to 
$k \alt 0.12\, h/\mathrm{Mpc}$ (for $z=0$). At first sight, this agreement
 could lead to the conclusion that non-linearities are small in this range 
 of wavenumbers. However, this is the effect of a fortuitous 
 cancellation between two non-linear effects, {\it i.e.} the damping of
 $P^I_{11}$ and the contribution $P^{II}_{11}$, which -- taken
  separately -- are of order 
 $40\%$ of the linear contribution.

\begin{figure}
\centerline{\includegraphics[width = 8in,
keepaspectratio=true,angle=-90
]{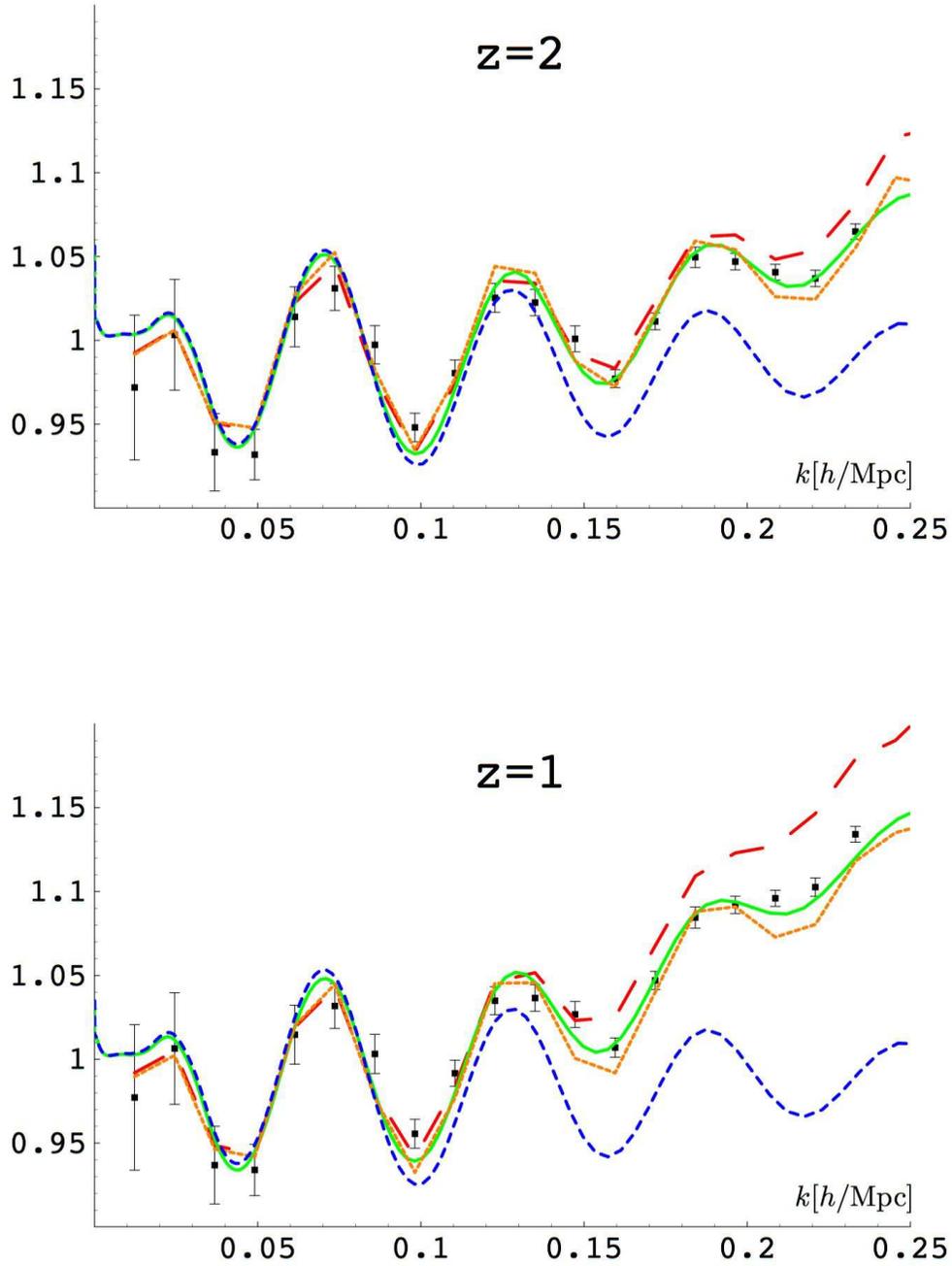}}
\caption{The power-spectrum at $z=2,\,1$, as given by the RG 
(solid line), linear theory (short-dashed), 1-loop PT
(long-dashed), the halo approach of Ref.~\cite{Smith} (very short-dashed), 
and the N-body simulations of \cite{Komatsu} (squares). The background cosmology is a spatially flat $\L$CDM model 
with $\Omega_\L^0=0.73$, $\Omega_b^0=0.043$, $h=0.7$, $n_s=1$}
\label{PJK}
\end{figure}

In Figs.~\ref{Pplot}, \ref{PJK} we compare our results (solid lines) 
with two sets of N-body simulations appeared in the literature, 
that is with those of refs.\cite{White} and \cite{Komatsu}, respectively. 
Since the two sets of simulations correspond to two slightly different 
cosmologies (see the figure captions), it is not possible to draw them 
on the same plot. The short-dashed lines correspond to linear theory and the 
long-dashed ones to 1-loop PT (which, at $z=0$ in Fig.~\ref{Pplot} 
has been truncated for $k\agt 0.17$ h/Mpc, where $P^{I}_{11}$ takes 
negative values, signaling the breakdown of the perturbative expansion). 
The black squares are taken from the N-body simulations. 
In Fig.\ref{PJK} we also plot the results from the halo-model approach of 
Ref.~\cite{Smith}. To enhance the BAO feature, each 
power-spectrum has been divided by the linear one, in a model without baryons
\cite{Eisenstein}.
In the peak region, our RG results agree with those of
N-body simulations to a few percent accuracy
down to redshift $z=0$, where linear and 1-loop
PT badly fail. Thus, the dynamical behavior in this
momentum range appears to be captured fairly well by the approximations 
implemented by our approach, namely the `single stream approximation', 
leading to Eq.~(\ref{compact}), and the non-linear corrections
of the two-point functions only, i.e. the propagator and the power-spectrum. 

An estimate of the accuracy and range of validity of the different 
approximations described so far can be made by improving the level 
of truncations, the first step being taking into account the running 
of the trilinear vertex. Alternatively, one can change the {\it ansatze} 
for the compositions between full propagators, Eq.~(\ref{compo}) or for
the time dependence of the $P^{II}$ contribution to the power-spectrum, 
Eqs.(\ref{ans1}, \ref{ans2}).
In Fig.~\ref{P2ansatz} we compare the results for the power-spectrum 
obtained by using the two different approximations for $P^{II}$. 
As one can can see, the results are quite stable in the range of 
momenta relevant
for the BAO, that is, for $k\alt 0.25\,\mathrm{h \,Mpc^{-1}}$. 
For higher momenta, the power-spectrum obtained from the RG is suppressed 
with respect to the one from the N-body simulations. The 
main origin of this effect is the progressive failure of the composition 
law of Eq.~(\ref{compo}) for large values of the combination 
$\sigma_v^2 k^2(e^{s_1}-e^{s_3})^2$.  
In order to improve this approximation, one could split into small 
steps the time integrations, that now are made in one single step from 
$\eta_i=\log D^+(0)/D^+(z_{in})$ down to $\eta=\log D^+(z)/D^+(z_{in})$, 
with $z_{in}$ typically larger than 30. At each step the renormalized 
power-spectrum would be assumed as the new `primordial' one. 
In this way, we expect to reduce considerably the dependence on the 
different aproximations to the exact time-dependence of the propagator 
and of $P^{II}$. The implementation of this procedure, together 
with the inclusion of the running of the vertex, will be the 
subject of future work.

\section{Conclusions}
In this paper we have explored the possibility of using RG techniques 
as a computational tool to solve the continuity and Euler 
equations (\ref{Euler}) non-perturbatively. 
The need of a non-perturbative
approach as one moves towards high wavenumbers is manifest if 
one considers the propagator. In this case, the $n$-th order perturbative 
contribution diverges as $k^{2n}$, while the full result goes exponentially 
to zero, as shown in Ref.~\cite{Crocce2} and in 
this paper. As a consequence, the full RG flow freezes at large momenta,
and the results are insensitive to the details of the matter distribution
at small scales, an effect which cannot be seen at any order in PT.

\begin{figure}
\centerline{\includegraphics[width = 6.5in,
keepaspectratio=true
]{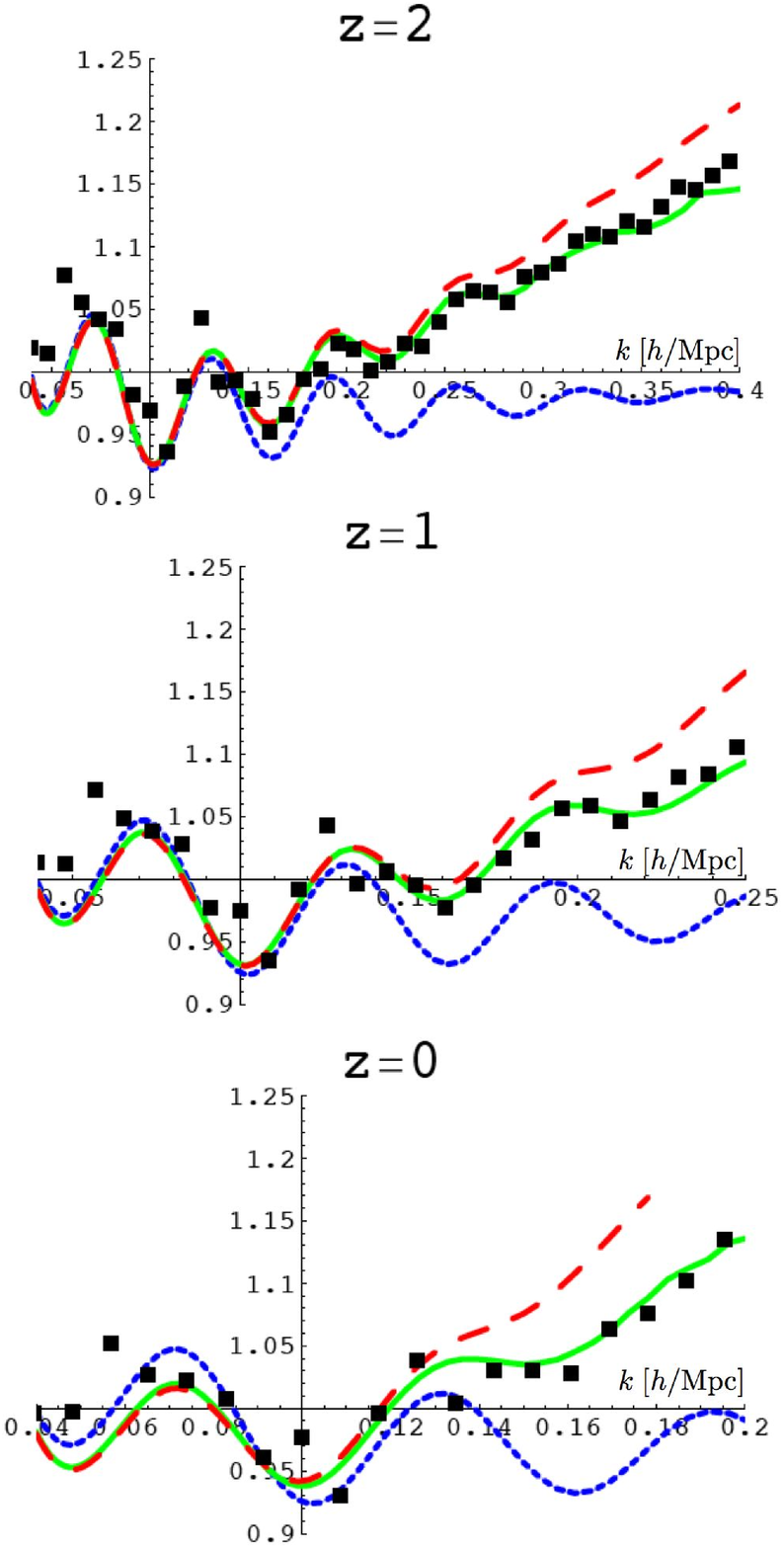}}
\caption{The power-spectrum at $z=2,\,1,\,0$, as given by the RG 
(solid line), linear theory (short-dashed), 1-loop PT
(long-dashed), 
and the N-body simulations of \cite{White} (squares). The background cosmology is a spatially flat $\L$CDM model 
with $\Omega_\L^0=0.7$, $\Omega_b^0=0.046$, $h=0.72$, $n_s=1$.}
\label{Pplot}
\end{figure}

Non-linear effects contribute sizeably to the power-spectrum in the BAO 
momentum range (see Fig.~\ref{P1P2}), and their reliable estimation is 
completely out of reach of PT, for $z\alt 1$ (Fig.\ref{Pplot}). We have
approximated the RG equations by renormalizing only the 2-point functions, 
{\it i.e.} the power-spectrum and the propagator,  
while keeping the trilinear interaction frozen at its tree-level value. 
This approximation performs surprisingly well, and our results 
agree with those of the N-body simulation of \cite{White}.

In view of other applications requiring the power-spectrum 
at higher wavenumbers, like {\it e.g.} weak gravitational lensing,  
the RG performance 
can be systematically improved by 
increasing the level of truncation of the full tower of differential 
equations. The next step will be the inclusion of the running of the 
trilinear vertex, which will also allow a computation 
of non-linear effects on the bispectrum.  

Further applications of our RG approach include the computation 
of higher-order statistics, the 
analysis of the possible effects of initial non-Gaussianity on 
non-linear scales and the effect of a non-vanishing stress-tensor 
$\sigma_{ij}$. 

The extension of the present approach to cosmological models
 including dynamical dark energy will also be the subject of future
 work.

\begin{figure}
\centerline{\includegraphics[width = 6.5in,
keepaspectratio=true,angle=0
]{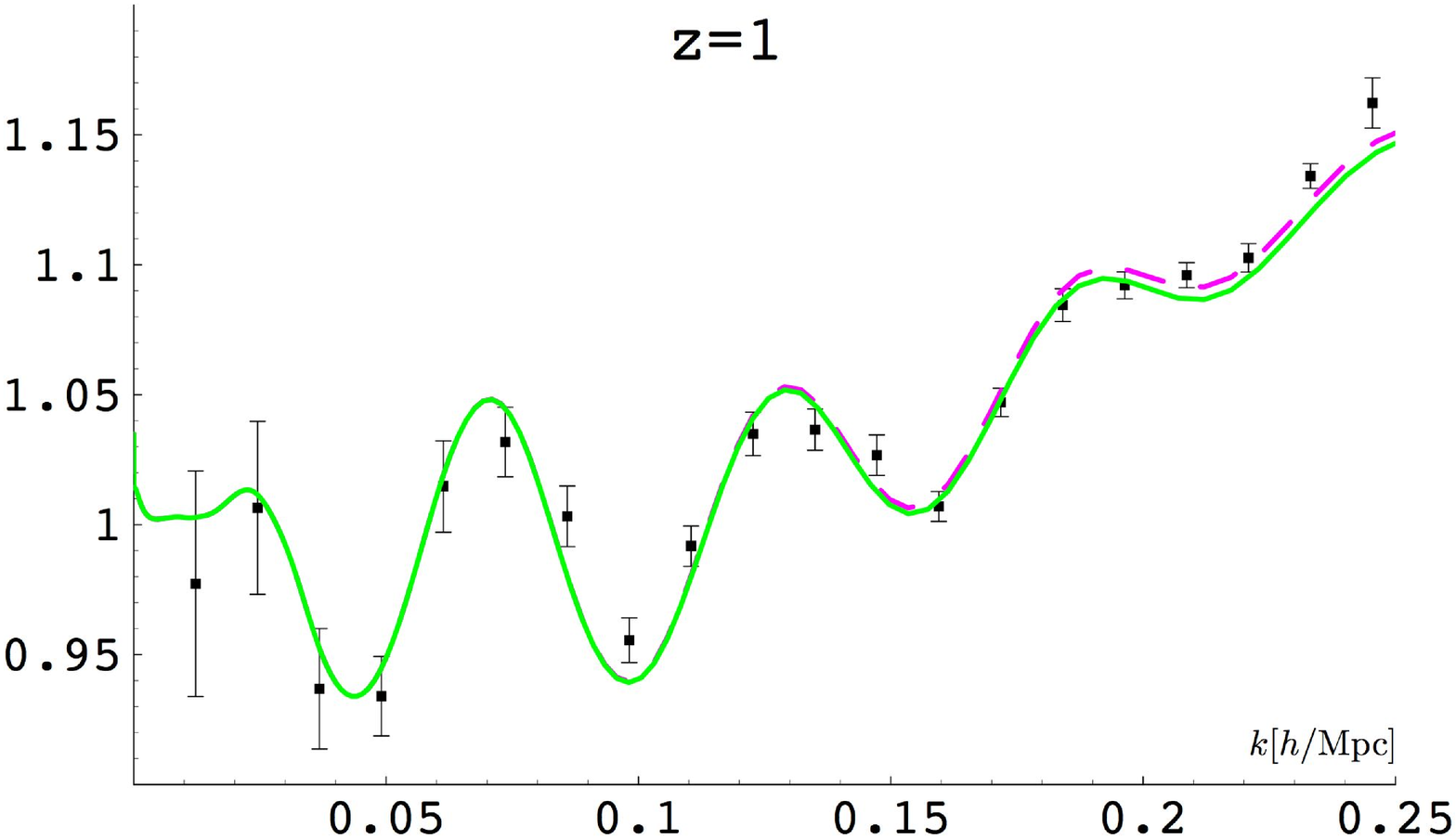}}
\caption{The power-spectrum at $z=1$, as given by the RG with two different {\it ansatze} for the time-dependence of $P^{II}_\l$. The
 dashed and solid lines were obtained using Eqs.~(\ref{ans1}) and (\ref{ans2}), respectively. Dots with error bars are taken from the simulations of Ref.~\cite{Komatsu}.}
\label{P2ansatz}
\end{figure}

\noindent
 \section*{Acknowledgments}
We thank E. Komatsu and M. White, for providing us with the N-body data of 
Refs.~\cite{Komatsu} and \cite{White} respectively, N. Bartolo, P. McDonald and M. Viel, for discussions. 
M.P. thanks the Galileo Galilei Institute for Theoretical Physics 
for hospitality during the initial stages of this work.

\end{document}